\documentclass[11pt]{article}
\usepackage{amsmath,amssymb,bm,epsf,epsfig,graphicx,hyperref,physics, cleveref, subfig}
\usepackage[backend=bibtex,natbib,style=numeric-comp,sorting=none, doi=false, isbn=false,url=false]{biblatex}
\input epsf.sty
\newcommand{\sectiono}[1]{\section{#1}\setcounter{equation}{0}}

\def\bra#1{\langle #1 |}
\def\ket#1{|#1 \rangle}

\def \be {\begin{equation}}
\def \ee {\end{equation}}
\def \bea {\begin{eqnarray}}
\def \eea {\end{eqnarray}}
\def \bdm {\begin{displaymath}}
\def \edm {\end{displaymath}}

\def \cc {{\mathbb C}}
\def \rr {{\mathbb R}}

\def \dd {{\cal D}}

\def\del {\partial}
\def\0{\nonumber}

\def\threept{C_{VVV}}

\begin{document}
\begin{center}
 {\large \bf $\,$\\
\vskip2cm
Closed string tachyon condensation revisited}

\vskip 1.1cm

{\large Jaroslav Scheinpflug\footnote{Email:
jaroslavscheinpflug at gmail.com}$^{(a)}$
,  Martin Schnabl\footnote{Email:
schnabl.martin at gmail.com}$^{(a)}$
} \vskip 1 cm

$^{(a)}${\it {Institute of Physics of the ASCR, v.v.i.} \\
{Na Slovance 2, 182 21 Prague 8, Czech Republic}}
\end{center}

\vspace*{6.0ex}

\centerline{\bf Abstract}
We consider condensation of nearly marginal matter tachyons in closed string field theory and observe that upon restricting to a subspace of states not containing the ghost dilaton, the on-shell value of the action is proportional to the shift of the central charge of the matter CFT. This correspondence lets us find a novel conformal perturbation theory formula for the next-to-leading order shift of the central charge for a generic theory, which we test on Zamolodchikov's flow between consecutive minimal models. Upon reintroduction of the dilaton couplings, it is plausible to have a vanishing value of the on-shell action.
\bigskip

\vfill \eject

\baselineskip=15pt

\tableofcontents

\sectiono{Introduction}
Closed string tachyon condensation is a subject with a long history \cite{Adams:2001sv, Dabholkar:2001if, Harvey:2001wm, Vafa:2001ra, Gregory:2003yb, Moeller:2003gg, Hellerman:2004qa, Hellerman:2006hf, Hellerman:2006ff,  Yang:2005rw, McGreevy:2005ci, Bergman:2006pd, Wang:2016sib} and due to the successes of open string field theory (OSFT) in the description of the open string analogue \cite{Sen:1999nx, Schnabl:2005gv, Erler:2007xt, Erler:2009uj, Kudrna:2014rya, Kudrna:2018mxa, Erler:2019xof, Kudrna:2021rzd, Scheinpflug:2023osi} (see \cite{Erler:2019vhl} for a recent review), it is natural to investigate tachyon condensation in closed string field theory (CSFT) \cite{Okawa:2004rh, Yang:2005rx, Moeller:2006cv, Moeller:2008tb, Schlechter:2019hrb} (see \cite{Zwiebach:1992ie,deLacroix:2017lif, Erler:2019loq,Erbin:2021smf, Maccaferri:2023vns} for reviews). This has been complicated by the notorious difficulty in explicitly constructing the CSFT vertices \cite{Zwiebach:1990ni, Zwiebach:1990nh, Moeller:2004yy, Moeller:2006cw, Moosavian:2017qsp, Moosavian:2017sev, Moosavian:2019zww,  Headrick:2018dlw, Headrick:2018ncs, Costello:2019fuh, Cho:2019anu, Firat:2021ukc, Ishibashi:2022qcz}, but in light of the recent advances \cite{Erbin:2022rgx, Firat:2023glo, Firat:2023suh} we might be moving to an era where the use of CSFT vertices becomes practical.

We revisit closed string tachyon condensation in CSFT by considering a setup where the tachyon is a nearly marginal spinless matter primary. Such a setting was considered previously in both the closed string \cite{Sen:1990hh,Mukherji:1991tb, GHOSHAL1991295} and the open string \cite{Scheinpflug:2023osi}. This setup enables us to bypass some of the difficulties associated with CSFT vertices and analytically study the properties of the resulting tachyon vacuum to quartic order. An important technical step is to use the infinite stub limit \cite{Sen:2019jpm, Schnabl:2023dbv, Erbin:2023hcs}, which lets us sidestep flattenization of the closed string propagator \cite{Berkovits:2003ny}. The resulting scheme is then a very efficient hybrid between the pure SFT scheme of \cite{Scheinpflug:2023osi} and point-splitting, which is the go-to regularisation scheme in conformal perturbation theory \cite{Ludwig:1987gs,Zamolodchikov:1987ti, Affleck:1991tk,Ludwig:1994nf, Gaberdiel:2008fn, Gaiotto:2012np, Poghossian:2013fda,Poghosyan:2013qta, Ahn:2014rua, Komargodski:2016auf, Konechny:2020jym, Poghosyan:2022ecv, Poghosyan:2023brb, Konechny:2023xvo}.

When we restrict ourselves to a subspace of states, which does not contain the ghost dilaton, we find a remarkable relation between the depth of the tachyon potential and the shift of the central charge of the matter CFT (we set $g_c = 1$)
\be
S = -\frac{1}{12} \Delta c,
\label{sko}
\ee
which is quite analogous to the relation between the on-shell OSFT action and the shift of the $g$-function $S = -\frac{1}{2\pi^2} \Delta g$ \cite{Sen:1999xm, Erler:2014eqa}. In contrast to the OSFT case, it is not clear to us what the target space interpretation of this result is. Note that the proportionality factor was fixed by comparison with the leading order result of \cite{Ludwig:1987gs}. Solving the CSFT equations of motion to quartic order gives
\be
\Delta c = -\frac{y^3}{\threept^2} - \frac{ \mathcal{A}_{TTTT}^{f.p.}}{2\threept^4} y^4 + O(y^5),
\label{cecko}
\ee
where $y = 2(1-h)$ is the RG eigenvalue for a perturbation by a $(h,h)$ spinless matter primary $V$ with fusion $V \cross V = 1 + V$ and $\mathcal{A}_{TTTT}^{f.p.}$ is the finite part (we regularised worldsheet UV divergences) of the appropriately regularised amplitude of four tachyons $T = t c_1 \bar{c}_1 V\ket{0}$
\bea
\label{fourtachyon}
\mathcal{A}_{TTTT}^{f.p.} &=& -\frac{1}{2\pi i} \int\displaylimits_\cc d\xi \wedge d\bar{\xi} \, \, \biggr[ \bra{V} V(1) V(\xi,\bar{\xi}) \ket{V} - \frac{1}{2}\biggr(\frac{1}{\abs{\xi}^4\abs{\xi-1}^4} + \frac{\abs{\xi-1}^4}{\abs{\xi}^4} + \frac{\abs{\xi}^4}{\abs{\xi-1}^4} \nonumber \\\ && \hspace{1 cm}-4 \biggr[\frac{1}{\abs{\xi}^2\abs{\xi-1}^2} + \frac{1}{\abs{\xi}^2} + \frac{1}{\abs{\xi-1}^2}\biggr]\biggr)  - \frac{\threept^2}{2}\biggr( \frac{1}{\abs{\xi}^2\abs{\xi-1}^2} + \frac{1}{\abs{\xi}^2} + \frac{1}{\abs{\xi-1}^2}\biggr) \biggr] \nonumber \\ && \, - 2 - \frac{4}{c}
\eea
with $c$ the central charge of the initial background and $\threept$ the tachyon structure constant. We check the validity of this formula by comparing with the exact answer for Zamolodchikov's flow between consecutive minimal models \cite{Zamolodchikov:1987ti}. Thus as emphasized in \cite{Larocca:2017pbo,Sen:2019jpm, Sen:2020cef, Sen:2020eck, Scheinpflug:2023osi}, string field theory naturally tames divergences on the worldsheet where conformal perturbation theory lives.

The result (\ref{sko}) is naively in tension with \cite{Erler:2022agw}, which states that the CSFT action vanishes on a solution. But one has to remember that this only holds when one includes the full state space of CSFT. In fact our solution encounters an obstruction in solving the equations of motion coming from the quartic couplings to the ghost dilaton. We find that for the result of \cite{Erler:2022agw} to hold, the ghost dilaton has to gain a nonperturbative VEV, a condition that we verify by studying the quartic couplings of dilatons to tachyons.

The paper is organised as follows. In section \ref{sec1} we calculate the quartic order nearly marginal tachyon potential for CSFT truncated to a subspace not containing the ghost dilaton and observe a relation between its depth and the shift of the matter central charge. In section \ref{sec2} we reintroduce the couplings to the ghost dilaton and observe that the ghost dilaton gains a nonperturbative VEV, providing a necessary condition for the on-shell value of the action to vanish. In (\ref{conclusion}) we conclude the paper. The appendix \ref{reg} containing the derivation of the formula (\ref{fourtachyon}) is supported by appendix \ref{lens}, which contains integrals over certain lens-like regions \cite{Poghossian:2013fda}. In appendix \ref{diff} we review how to extract local coordinates from quadratic differentials for the four punctured sphere case (see for example \cite{Moeller:2004yy}) and in \ref{integrands} we present the derivation \cite{Yang:2005ep} of the relevant integrands present in the couplings of dilatons to tachyons in \ref{sec2}, both for reader's convenience. Lastly, we present appendix \ref{osft} which is an introduction to some of the large stub techniques in the simpler setting of OSFT.

\sectiono{The tachyon potential and the central charge}
\label{sec1}
In this section we observe the relation between the depth of the tachyon potential calculated on a subspace of states not containing the ghost dilaton and the shift of the central charge of the matter CFT in which tachyon condensation happens. We note that appendix \ref{osft} serves as a warm-up to some of the technical aspects of our analysis.
	\subsection{Choice of vertex}
	\label{choicev}
	Consistent vertices of closed string field theory are notoriously difficult to explicitly construct. In this work, we rely on the observation of Sen \cite{Sen:2019jpm} that many quantities of interest can be calculated without the explicit knowledge of such vertices.

	In string field theory, we decompose the string moduli space (for us the moduli space of an $n$-punctured sphere $\Sigma_{0,n}$) into vertex and Feynman regions, where the latter covers regions near string degeneration. The Feynman region is interpreted as arising from surfaces that are obtained by gluing of surfaces with fewer punctures with a propagator. The consistency of this gluing imposes certain restrictions on the string vertices as embodied in the geometric BV equation \cite{Sen:1994kx, Maccaferri:2023vns}. Since punctures introduce sinks of curvature, they lead to unphysical metric dependence \cite{Erbin:2021smf}. To remedy this, we cut out parts of the $\Sigma_{0,n}$ around the punctures and replace them with flat discs. This means that we have a local relation between the coordinates $w$ of these discs and the global coordinate $z$ on $\Sigma_{0,n}$. We write this as
	\be
	w_i = h_i(z),
	\ee
	where $i \in {1,\ldots,n}$ labels the punctures located at $z_i$. We usually take a symmetric vertex for which the maps $h_i$ are related to one another as to ensure invariance under the permutation of the punctures.

\begin{figure}[h!]
	    \centering
    \subfloat[\centering Unstubbed polyhedral vertex region]{{\includegraphics[width=8cm]{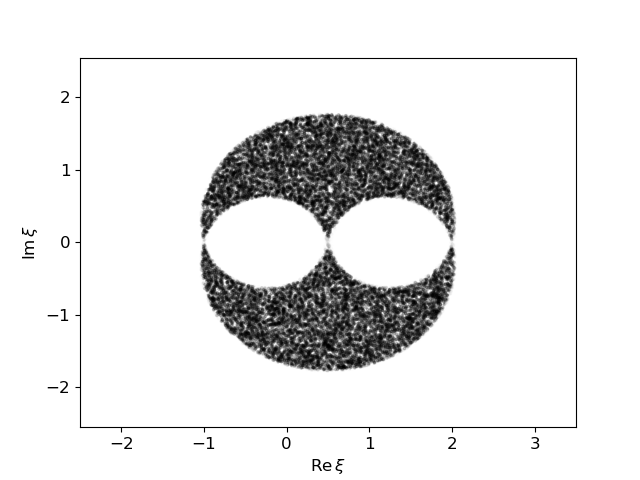} }}%
    \quad
    \subfloat[\centering Stubbed vertex region]{{\includegraphics[width=8cm]{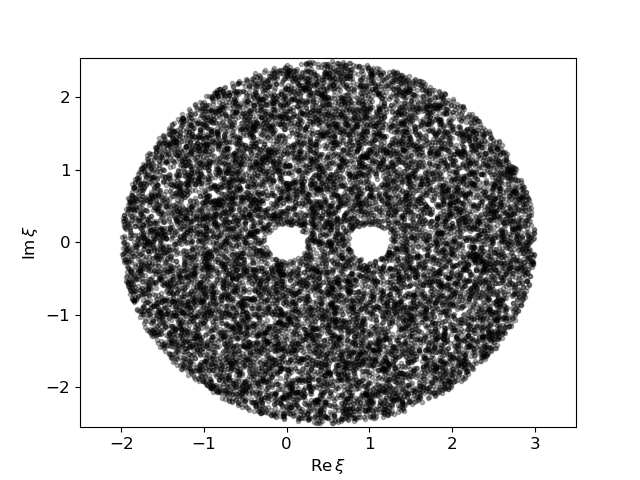} }}%
    \caption{Unstubbed vs. stubbed vertex region (in black), left figure drawn with the help of the AI of \cite{Erbin:2022rgx}, which is available in open source at \url{https://github.com/HaroldErbin/pysft-nn}}
	    \label{vertex}
\end{figure}

	In our work it will be very useful to make the vertex region as large as possible, since then the Feynman regions shrink to points and we can effectively bypass flattenization \cite{Berkovits:2003ny} of the string propagator. Another reason for doing this is that the shape of the vertex region becomes simpler, see figure \ref{vertex}, where we draw an actual polyhedral vertex region for the quartic vertex and illustrate the universal large stub behavior (not an actual stubbed vertex region is drawn, just $\cc$ with excisions). We make the vertex region larger by introducing large stubs \cite{Schnabl:2023dbv, Erbin:2023hcs}, which means that we scale the local coordinates by large $\lambda$ so that
	\be
	w_i = \lambda h_i(z) \,.
	\ee
Note that since we expect observables not to depend on the precise details of the implementation of the $h_i$ maps (as long as they produce non-overlapping discs), we also expect the $\lambda$ dependence to disappear and this gives a nice check on the validity of our computations.

Let us now concentrate on determining the shape of the stubbed quartic vertex, which covers a large part of the $\Sigma_{0,4}$ moduli space, which is $\Sigma_{0,3}$. We determine it as the complement of the Feynman region. On the three-punctured sphere $\Sigma_{0,3}$ we use $Sl(2,\cc)$ maps to map three of our punctures to $z_1 = 0$, $z_2 = \infty$, $z_3 = 1$. To obtain the Feynman region of $\Sigma_{0,4}$, we glue two $\Sigma_{0,3}$ by a propagator and write the global coordinates on these three-punctured spheres as $z$ and $z'$.
	By sewing at the third puncture as appropriate for the $s$-channel Feynman region $\mathcal{R}_s$ around 1 (the $u$ and $t$ channel Feynman regions $\mathcal{R}_u$ and $\mathcal{R}_t$ live near $0$ and $\infty$ respectively), we have
	\be
	w_3 w'_3 = \lambda^2 h_3(z) h_3(z') = q
	\label{sewing}
	\ee
	with $q$ the propagator modulus with $\abs{q} \leq 1$. This results in a four-punctured sphere with a new global coordinate. We apply an $Sl(2,\cc)$ map to move $z = 0$, $z = \infty$ and $z' = \infty$ to the three canonical points 0, $\infty$ and $1$. The image of $z' = 0$ under this $Sl(2,\cc)$ map we then call $\xi$ and it is the moving puncture, which moves around the moduli space $\Sigma_{0,3}$. We now ask how to relate $\xi$ to the Feynman region modulus $q$. To do this, expand (\ref{sewing}) around $z=1$ and $z'=1$
	\be
	\lambda^2 \big(h'_3(1)\big)^2(z-1)(z'-1) + \ldots = q
	\ee
	so that $z'=0$ corresponds to
	$z = 1 - \frac{q}{\big(\lambda h'_3(1)\big)^2} + \ldots$. In the Feynman region, we can take the global coordinate to be $z$, so that
	\be
	\xi = 1 - \frac{q}{\big(\lambda h'_3(1)\big)^2}+ \ldots
	\label{gSChan}
	\ee
	We have now related $\xi$ and $q$ in the $s$-channel and we see that in the large $\lambda$ limit the boundary of the $s$-channel Feynman region $\abs{q}=1$ becomes a small circle around 1 in the $\xi$ plane. Analogously in the $u$-channel, we have a circle around 0 of the same radius and in the $t$-channel a large circle around 0 of radius $\abs{\lambda h'_3(1)}^2$ (which is a small radius around $\infty$). This gives the universal shape of the largely stubbed quartic vertex in figure \ref{vertex}.

\subsection{Solving the equations of motion}

Our CSFT background consists of a matter CFT, which is given by a product of a CFT of interest with central charge $c$ and a spectator CFT with central charge $26-c$, and the ghost CFT. Since we want to go beyond leading order, a key assumption is that the matter CFT of our interest is not an isolated CFT, but lives on some manifold of CFTs along which we know the CFT data such as $\threept$, see \cite{Komargodski:2016auf}. Further, on such a manifold there must exist an accumulation point of CFTs with a nearly marginal primary operator $V$ of weight $\Delta$ in their spectrum such that at this point this operator becomes marginal with $y\equiv 2 - \Delta =0$. Our analysis then holds for this family of nearly marginal operators that is parametrically connected to $y=0$. Usually we deform by a nearly marginal member of this family with very small $y$, but if we push our perturbation expansion to high enough order (convergence permitting), this can be relaxed. If we were interested only in the leading order result, we could simply pretend our deforming operator is a part of such a family \cite{Komargodski:2016auf}.

After formulating our theory on a background from such a manifold, we perturbatively solve the CSFT equations of motion
\be
Q \Phi + \frac{1}{2!} [\Phi,\Phi] + \frac{1}{3!} [\Phi,\Phi,\Phi] + \ldots = 0
\ee
that follow from the CSFT action \cite{Zwiebach:1992ie} (for reviews see the references mentioned in the introduction)
\be
S = \frac{1}{2!} \langle \Phi, Q\Phi \rangle + \frac{1}{3!} \langle \Phi,\Phi,\Phi\rangle + \frac{1}{4!} \langle \Phi,\Phi,\Phi,\Phi \rangle + \ldots
\label{action}
\ee
where we write $\langle A_0, A_1, A_2, \ldots \rangle \equiv \braket{A_0 }{c_0^{-}[A_1,A_2,\ldots]}$ and set $g_c = 1$.
As usual \cite{Erbin:2020eyc} we consider the ansatz $\Phi = T + X$, where $T = t c_1\bar{c}_1 V\ket{0}$ is the tachyon with $V$ a member of the family of nearly marginal matter primaries mentioned in the previous paragraph. Further, we specialize to the spinless case of an $(h,h)$ operator so that $y = 2(1-h)$ with fusion $V \cross V = 1 + V$. The $X = \bar{P} \Phi$ is the contribution from the fields that are integrated out, where $\bar{P} = 1 - P$ with $P$ a projector onto $T$ or $Q T$ (depending on the ghost number).

We project the equations of motion onto $T$ and $X$ respectively
\bea
\label{firstEOM} Q T + \frac{1}{2!} P[\Phi,\Phi] + \frac{1}{3!} P [\Phi,\Phi,\Phi] + \ldots &=& 0 \\
\label{secondEOM} Q X + \frac{1}{2!} \bar{P}[\Phi,\Phi] + \frac{1}{3!} \bar{P} [\Phi,\Phi,\Phi] + \ldots &=& 0
\eea
and solve (\ref{secondEOM}) in Siegel gauge $b_0^+ \Phi = 0$
\be
X = H \biggr(\frac{1}{2!} [\Phi,\Phi] + \frac{1}{3!} [\Phi,\Phi,\Phi] + \ldots \biggr)
\label{XSol}
\ee
with $H = - \frac{b_0^+}{L_0^+} \bar{P}$ being the Siegel gauge homotopy operator. At this point, we need to restrict ourselves to a subspace of the full state space not containing the ghost dilaton since $\langle D,T,T,T\rangle \neq 0$ with $D = d (c_1 c_{-1} - \bar{c}_1\bar{c}_{-1})\ket{0}$ being a nontrivial element of the cohomology. This means that there is a dilaton component in (\ref{secondEOM}) coming from $\bar{P} [\Phi,\Phi,\Phi]$, which we cannot cancel by tuning $X$ since $Q X$ is exact. The equation (\ref{firstEOM}) is an algebraic equation for the fixed point value of the coupling $t$ and by projecting, we write it as
\bea
\langle T, Q T \rangle + \frac{1}{2!} \langle T, \Phi,\Phi \rangle + \frac{1}{3!} \langle T, \Phi, \Phi, \Phi \rangle + \ldots &=& 0 \,.
\eea
Plugging in $\Phi = T + X$ with $X$ from (\ref{XSol}), we have
\be
\begin{aligned}
    &\langle T, Q T \rangle + \frac{1}{2!} \langle T, T,T \rangle + \frac{1}{3!} \langle T, T, T, T \rangle + \frac{1}{2!} \langle T,T, H[T,T] \rangle + \ldots = 0 \,.
\end{aligned}
\ee
which simplifies by noting that the last two terms can be combined into the nearly on-shell (zero-momentum) four-point amplitude $t^4 \mathcal{A}_{TTTT} \equiv \langle T,T,T,T \rangle + 3 \langle T,T,H[T,T]\rangle$
\be
\begin{aligned}
    &\langle T, Q T \rangle + \frac{1}{2!}\langle T,T,T\rangle + \frac{1}{3!} t^4 \mathcal{A}_{TTTT} + \ldots &= 0 \label{tachyonEqq}
\end{aligned}
\ee

To evaluate the fixed point coupling $t$ to leading order, we need to compute $\langle T, Q T \rangle$ and $\langle T,T,T\rangle$. We have
\bea
\langle T, Q T \rangle &=& t^2 \braket{c_{-1}\bar{c}_{-1} V }{\frac{1}{2}(c_0-\bar{c}_0)(c_0 L_0 + \bar{c}_0 \bar{L}_0) c_1 \bar{c_1} V } \nonumber \\ &=& \frac{t^2 (h-1)}{2}\biggr( \braket{c_{-1}\bar{c}_{-1}  }{c_0\bar{c}_0 c_1 \bar{c_1}  } -  \braket{c_{-1}\bar{c}_{-1} }{\bar{c}_0c_0 c_1 \bar{c_1}  }\biggr) \nonumber \\ &=& - t^2 (h-1) \nonumber \\&=& \label{kinetic} \frac{t^2 y}{2}
\eea
where we used that the BPZ conjugate of $c_1 \bar{c}_1 \ket{0}$ is $\bra{0}c_{-1} \bar{c}_{-1}$ and that we normalise the ghosts as $\bra{0} c_{-1}\bar{c}_{-1} c_0 \bar{c}_0c_1\bar{c}_1\ket{0} = -1$ and matter as $\braket{V}{V} = 1$. The three tachyon coupling can be written using the local coordinate maps as
\be
\langle T,T,T\rangle = -t^3 \threept \abs{\lambda h'_3(1)}^{3y} = -t^3 \threept (1 + 3y \ln \abs{\lambda h'_3(1)} + \ldots)
\ee
where we used the permutation symmetry of the vertex and got a minus from ghosts and $\threept$ from matter. Plugging these into (\ref{tachyonEqq}) truncated to cubic order, we get
\be
\frac{t^2 y}{2} - \frac{1}{2!} t^3 \threept + \ldots = 0
\ee
so that
\be
\label{fpoint}
t = \frac{y}{\threept} + O(y^2) \,.
\ee
We see that a perturbation expansion in $t$ is equivalent to a perturbation expansion in $y$, which is small for $V$ nearly marginal.

Next we would like to compute the depth of the tachyon potential (on-shell value of the action) at the fixed point and to do this we note that thanks to the equations of motion it simplifies to
\be
S = \frac{1}{6}\biggr(\langle T,Q T\rangle - \frac{1}{12} t^4 \mathcal{A}_{TTTT}\biggr)+ \ldots \, .
\label{actionOS}
\ee
Plugging in (\ref{kinetic}) with the fixed point coupling (\ref{fpoint}), we have to cubic order
\be
S = \frac{1}{6} \frac{t^2 y}{2} + \ldots = \frac{1}{12} \frac{y^3}{\threept^2} + O(y^4)
\ee
in apparent contradiction with \cite{Erler:2022agw} but this is understandable since we are omitting the ghost dilaton. Note that in our normalisation we get to leading order $S = -\frac{1}{12} \Delta c$ with $\Delta c = -\frac{y^3}{\threept^2} + O(y^4)$ being the shift in the matter CFT central charge known from conformal perturbation theory \cite{Ludwig:1987gs}. A natural question is whether this relation persists to quartic order and this is what we now turn to.

We need to calculate the amplitude of four nearly on-shell tachyons $\mathcal{A}_{TTTT}$. To do this, we start with the Feynman region contribution
\be
3 \langle T,T,H[T,T]\rangle = - 3 \langle T,T, \frac{b_0^+}{L_0^+} \bar{P} [T,T]\rangle = - 3 \langle T,T, \frac{b_0^+}{L_0^+} \bar{P} b_0^- c_0^-  [T,T]\rangle
\ee
where we used that the closed string product output is annihilated by $b_0^-$ since twisting is integrated out. We now insert the identity \cite{Sen:2019jpm} in the untwisted Hilbert space $\sum_i \ket{\xi_i}\bra{\xi^c_i} = \ket{\xi^c_i}\bra{\xi_i} = 1 $ with $b_0^- \ket{\xi_i} = L_0^- \ket{\xi_i} = 0$ and $\braket{\xi^c_i}{\xi_j} = \braket{\xi_i}{\xi^c_j} = \delta_{i j}$ to obtain
\be
- 3 \langle T,T, \frac{b_0^+}{L_0^+} \bar{P} b_0^- c_0^-  [T,T]\rangle = -3 \sum_{i j} \langle T,T, \xi_i\rangle \bra{\xi^c_i} \frac{b_0^+}{L_0^+} \bar{P} b_0^- \delta_{L_0^-} \ket{\xi^c_j} \langle \xi_j, T,T\rangle \, .
\ee
From the three-vertices, we get factors of $\abs{\lambda h'_3(1)}^{2y - \Delta_i}$ with $\Delta_i$ being the dimension of $\ket{\xi_i}$. Since we take $\lambda$ large, we see that states with $\Delta_i > 0$ are highly suppressed and noting that we take the fusion $V \cross V = 1 + V$ and that we have a projector $\bar{P}$, the only state with $\Delta_i \leq 0$ remaining (apart from the ghost dilaton) is the identity $\ket{\xi_i} = c_1\bar{c}_1 \ket{0}$ with $\Delta_i = - 2$ and the conjugate $\ket{\xi^c_i} = - c_0\bar{c}_0 c_1\bar{c}_1\ket{0}$. We thus get
\bea
3 \langle T,T,H[T,T]\rangle &=& -3 t^4 \abs{\lambda h'_3(1)}^{4y + 4} \bra{c_0\bar{c}_0c_1\bar{c}_1} \frac{b_0^+}{L_0^+} \bar{P} b_0^- \delta_{L_0^-} \ket{c_0\bar{c}_0 c_1\bar{c}_1} \\
                            &=& -3 t^4 \abs{\lambda h'_3(1)}^{4} + O(y)
\eea
where we used $\bra{c_0\bar{c}_0c_1\bar{c}_1} \frac{b_0^+}{L_0^+} \bar{P} b_0^- \delta_{L_0^-} \ket{c_0\bar{c}_0 c_1\bar{c}_1} = 1$.

The next step is to calculate the elementary four-tachyon coupling $\langle T,T,T,T \rangle$. To do this, we note that (see (\ref{integrands}))
\be
\langle T,T,T,T\rangle = -\frac{1}{2\pi i} t^4 \int\displaylimits_\mathcal{R} d \xi \wedge d \bar{\xi} \, \, \bra{V} V(1) V(\xi,\bar{\xi}) \ket{V} + O(y)
\ee
with $\mathcal{R}$ being the vertex region and we used that since we are nearly on-shell, we can discard the local coordinate maps to leading order. We now estimate the degree of divergence of this contact interaction in the limit $\lambda \to \infty$. To do so, we expand into conformal blocks \cite{DiFrancesco:1997nk} by noting the OPE structure of $V$s
\be
\bra{V} V(1) V(\xi,\bar{\xi}) \ket{V} = (\xi^{-2} + \ldots)(\bar{\xi}^{-2} + \ldots) + \threept^2 (\xi^{-1} + \ldots)(\bar{\xi}^{-1} + \ldots) + O(y)
\ee
where we expanded around $h = 1$ while keeping only the leading terms (the rest does not contribute to the divergences). In the large stub limit, the moduli space looks like a disc (in the complex plane) around zero of radius $\abs{\lambda h'_3(1)}^2$  with two small discs of radii $\abs{\lambda h'_3(1)}^{-2}$ cut out around zero and one, see figure \ref{vertex}. Now imagine a small annulus around zero of radii $\abs{\lambda h'_3(1)}^{-2}$ and $r_0$ with $r_0 < 1$ finite, which we integrate over and the dependence on the smaller radius gives the divergent contribution.  We write $\xi = x + i y$ so that $d \xi \wedge d \bar{\xi} = -2i r dr \wedge d\varphi$, $\frac{1}{\xi^2\overline{\xi}^2} = \frac{1}{r^4}$ and $\frac{1}{\xi \overline{\xi}} = \frac{1}{r^2}$.
\bea
-\frac{1}{2\pi i} (-2i) 2\pi \int\displaylimits_{\abs{\lambda h'_3(1)}^{-2}}^{r_0} dr\biggr(\frac{1}{r^3} + \threept^2 \frac{1}{r}\biggr) \sim \abs{\lambda h'_3(1)}^{4} + 4\threept^2 \ln\abs{\lambda h'_3(1)}
\eea
where we wrote only the divergent contribution.
Since there are three symmetrical regions of moduli space which produce this result to leading order in the stub expansion, one has
\bea
\mathcal{A}_{TTTT} &=& -3 \abs{\lambda h'_3(1)}^{4} + 3 \abs{\lambda h'_3(1)}^{4} + 12 \threept^2 \ln\abs{\lambda h'_3(1)} + \mathcal{A}_{TTTT}^{f.p.} + O(y) \nonumber \\
		   &=& 12 \threept^2 \ln\abs{\lambda h'_3(1)} + \mathcal{A}_{TTTT}^{f.p.} + O(y) \label{amp} \, .
		   \eea
		   Where the divergent contribution of the identity canceled between the Feynman and the vertex region.  The formula for $\mathcal{A}_{TTTT}^{f.p.}$ is derived in appendix \ref{reg}.

		   What we have done is essentially implementing a particular unambiguous point-splitting regularisation scheme where the small cutoff is played by $\abs{\lambda h'_3(1)}^{-2}$. A key feature of this scheme is that we don't have to worry about what happens inside the Feynman regions, since there divergences are handled by properly dividing by the eigenvalue of $L_0^+$. String field theory also automatically gave us a natural renormalised coupling $t$ so that we don't have to start by expanding in the bare coupling entering the action as has so far been the case in conformal perturbation theory.

		In order to compute the fixed point coupling to next-to-leading order, we plug (\ref{amp}) into (\ref{tachyonEq})
\be
\frac{y}{2} - \threept \frac{t}{2!} (1 + 3y \ln \abs{\lambda h'_3(1)}) + t^2 \frac{1}{3!} \biggr(\mathcal{A}_{TTTT}^{f.p.} + 12 \threept^2 \ln \abs{\lambda h'_3(1)} \biggr)+ O(y^2) = 0
\label{kapling}
\ee
which gives
\be
t = \frac{y}{\threept} + \frac{ \mathcal{A}_{TTTT}^{f.p.} + 3\threept^2 \ln \abs{\lambda h'_3(1)}}{3 \threept^3} y^2 + O(y^3) \,.
\ee
We see that in the large stub limit, the tachyon vacuum moves far away (at $y$ fixed) signaling that nonperturbative physics is obscured by stubs \cite{Erbin:2023hcs}.
The on-shell action (\ref{actionOS}) evaluated with this coupling is then
\be
S = \frac{1}{12} \frac{y^3}{\threept^2} + \frac{ \mathcal{A}_{TTTT}^{f.p.}}{24\threept^4} y^4 + O(y^5)
\ee
which is independent of the local coordinates and by $\Delta c = -12 S$ gives (\ref{cecko}). In other words the divergence coming from the $V$-channel in $\mathcal{A}_{TTTT}$ canceled with the divergence in $\langle T,T,T\rangle$. For us this result was a major clue to the fact that maybe the subsector of states which does not contain the dilaton contains interesting physics since one would perhaps expect some residual local coordinate dependence to be canceled by the couplings to the dilaton. We remark that although (\ref{cecko}) is a two loop result from the point of view of conformal perturbation theory, expressing everything in terms of amplitudes automatically removes one moduli integral with the same phenomenon happening in the open string, see (\ref{gres}).

Note that our computation can be carried out even in the case $\threept = 0$, but then we have to assume that $\mathcal{A}_{TTTT}^{f.p.} < 0$ (for $V$ relevant) in order for (\ref{kapling}) to find a pair of tachyon vacua at $t = \pm \sqrt{-\frac{y}{\mathcal{A}_{TTTT}^{f.p.}}}$. Both of these vacua give $S = -\frac{y^2}{8 \mathcal{A}_{TTTT}^{f.p.}} + O(y^\frac{3}{2})$ so that $\Delta c = \frac{3}{2} \frac{y^2}{\mathcal{A}_{TTTT}^{f.p.}} + O(y^\frac{3}{2})$.

\subsection{Testing the relation between the tachyon potential and the central charge}

We now want test the formula (\ref{cecko}).
The trivial test is in the case of the $c=1$ free boson, where
\be
\bra{V} V(1) V(\xi,\bar{\xi}) \ket{V} = \, \biggr(\frac{1}{\xi^2}+\frac{1}{(\xi-1)^2} + 1\biggr)\biggr(\frac{1}{\bar{\xi}^2} + \frac{1}{(\bar{\xi}-1)^2} + 1\biggr)
\ee
so that (\ref{fourtachyon}) becomes
\bea
\mathcal{A}_{TTTT}^{f.p.} &=& -\frac{1}{2\pi i} \int\displaylimits_\cc d\xi \wedge d\bar{\xi} \, \, \biggr[\biggr(\frac{1}{\xi^2}+\frac{1}{(\xi-1)^2} + 1\biggr)\biggr(\frac{1}{\bar{\xi}^2} + \frac{1}{(\bar{\xi}-1)^2} + 1\biggr) \nonumber \\ && \hspace{1cm}- \frac{1}{2}\biggr(\frac{1}{\abs{\xi}^4\abs{\xi-1}^4} + \frac{\abs{\xi-1}^4}{\abs{\xi}^4} + \frac{\abs{\xi}^4}{\abs{\xi-1}^4} -4 \biggr[\frac{1}{\abs{\xi}^2\abs{\xi-1}^2} + \frac{1}{\abs{\xi}^2} + \frac{1}{\abs{\xi-1}^2}\biggr]\biggr)  \nonumber\biggr] \\ && - 2 - 4 \,.
\eea
Going to polar coordinates by $\xi = r e^{i \varphi}$ and performing the angular integration, we get
\be
\mathcal{A}_{TTTT}^{f.p.} = 6\int\displaylimits_0^1 d r \,r + 6 \int\displaylimits_1^\infty \frac{d r}{r^3} \, - 2 - 4 = 0
\ee
as one would expect since exactly marginal deformations do not change the central charge.

The nontrivial test is for the initial background of choice being the unitary minimal model $\mathcal{M}_m$ with central charge $c_m = 1 - \frac{6}{m(m+1)}$ in the large $m$ limit. In this limit there exists an operator $V_{(1,3)} \equiv V$ with the fusion $V \cross V = 1 + V$ and dimension $2(1-\frac{2}{m}) \equiv 2(1-\epsilon) = 2 - y$ so that it is slightly relevant for $m$ large. The deformation by $V$ gives a canonical example of conformal perturbation theory in the bulk and was first studied by Zamolodchikov \cite{Zamolodchikov:1987ti} and later expanded upon for example by Gaiotto \cite{Gaiotto:2012np} and Poghossian \cite{Poghossian:2013fda}. Under this deformation one has the flow $\mathcal{M}_m \to \mathcal{M}_{m-1}$ so that the expected answer for $\Delta c$ is
\be
\Delta c = c_{m-1} - c_m = -\frac{12}{m(m^2-1)} = -\frac{3\epsilon^3}{(2-\epsilon)(1-\epsilon)} = - \frac{3}{2}\epsilon^3 - \frac{9}{4} \epsilon^4 + \ldots \,.
\ee
The relevant structure constant $\threept$ has an expansion \cite{Zamolodchikov:1987ti}
\be
\threept = \frac{4}{\sqrt{3}}(1-\frac{3}{2} \epsilon + \ldots)
\label{structure}
\ee
and by using it in (\ref{cecko}), we have the prediction
\be
\Delta c = -\frac{3}{2}\epsilon^3 - \frac{9}{32}(\mathcal{A}_{TTTT}^{f.p.} + 16)\epsilon^4 + \ldots \,.
\ee
Which means that if one would have $S = -\frac{1}{12} \Delta c$, then for this deformation $\mathcal{A}_{TTTT}^{f.p.} = -8 + O(\epsilon)$. We now show that this is exactly what happens. The relevant correlator is \cite{Poghossian:2013fda}
\be
\bra{V} V(1) V(\xi,\bar{\xi}) \ket{V} = \abs{\frac{1-2\xi+3\xi^2-2\xi^3+\frac{\xi^4}{3}}{\xi^2(1-\xi)^2}}^2 + \frac{16}{3}\abs{\frac{1-\frac{3\xi}{2}+\xi^2-\frac{\xi^3}{4}}{\xi(1-\xi)^2}}^2 + \frac{5}{9}\abs{\frac{\xi}{1-\xi}}^2 + O(\epsilon)
\ee
which after plugging into (\ref{fourtachyon}), using the expansion (\ref{structure}) and that $c=1$ in the large $m$ limit gives
\bea
\mathcal{A}_{TTTT}^{f.p.} &=& -\frac{1}{2\pi i} \int\displaylimits_\cc d\xi \wedge d\bar{\xi} \, \,\biggr[\abs{\frac{1-2\xi+3\xi^2-2\xi^3+\frac{\xi^4}{3}}{\xi^2(1-\xi)^2}}^2 + \frac{16}{3}\abs{\frac{1-\frac{3\xi}{2}+\xi^2-\frac{\xi^3}{4}}{\xi(1-\xi)^2}}^2 + \frac{5}{9}\abs{\frac{\xi}{1-\xi}}^2 \nonumber \\ && \hspace{1 cm}- \frac{1}{2}\biggr(\frac{1}{\abs{\xi}^4\abs{\xi-1}^4} + \frac{\abs{\xi-1}^4}{\abs{\xi}^4} + \frac{\abs{\xi}^4}{\abs{\xi-1}^4} -4 \biggr[\frac{1}{\abs{\xi}^2\abs{\xi-1}^2} + \frac{1}{\abs{\xi}^2} + \frac{1}{\abs{\xi-1}^2}\biggr]\biggr)  \nonumber \\ && \hspace{1 cm}- \frac{8}{3}\biggr( \frac{1}{\abs{\xi}^2\abs{\xi-1}^2} + \frac{1}{\abs{\xi}^2} + \frac{1}{\abs{\xi-1}^2}\biggr) \biggr] \nonumber \\ && - 2 - 4 + O(\epsilon) \nonumber \, .
\eea
Upon angular integration it becomes
\be
\mathcal{A}_{TTTT}^{f.p.} = -2\int\displaylimits_0^1 d r \,r -  2 \int\displaylimits_1^\infty \frac{d r}{r^3} \, - 2 - 4 + O(\epsilon) = -8 + O(\epsilon),
\ee
which confirms our proposal.

\sectiono{Adding the ghost dilaton}
\label{sec2}
In this section we repeat the analysis of section \ref{sec1} with the ghost dilaton $D = d(c_1 c_{-1} - \bar{c}_1 \bar{c}_{-1})\ket{0}$ added to our state space and show evidence for the finite rescaling of the tachyon vacuum of section \ref{sec1}. This avoids an immidiate contradiction with the result of Erler \cite{Erler:2022agw}. Thanks to the dilaton theorem to leading order in $y$ there is a noteworthy relation between the Feynman and vertex regions of amplitudes involving dilatons in the form of localisation, we which exploit in subsection \ref{elementary}.
\subsection{The equations of motion}
Enlarging our projector $P$ so that it projects onto $R \equiv T + D$, the equation $(\ref{firstEOM})$ becomes two algebraic equations for the fixed point values of the couplings $t$ and $d$ and by projecting, we write them as
\bea
\langle T, Q R \rangle + \frac{1}{2!} \langle T, \Phi,\Phi \rangle + \frac{1}{3!} \langle T, \Phi, \Phi, \Phi \rangle + \ldots &=& 0 \\
\langle D, Q R \rangle + \frac{1}{2!} \langle D, \Phi,\Phi \rangle + \frac{1}{3!} \langle D, \Phi, \Phi, \Phi \rangle + \ldots &=& 0 \,.
\eea
Using the fact that only the four-point couplings with $D$ are nonzero and $\langle V\rangle = 0$ while plugging in $\Phi = R + X$ with $X$ from (\ref{XSol}), we have
\be
\begin{aligned}
    &\langle T, Q T \rangle + \frac{1}{2!} \langle T, T,T \rangle + \frac{1}{3!} \langle T, T, T, T \rangle + \frac{1}{2!} \langle T,T, H[T,T] \rangle + \\  &\frac{1}{2!} \langle T,T,T,D\rangle + \langle T,T,H[T,D]\rangle+ \frac{1}{2!} \langle T,D,H[T,T]\rangle + \\& \frac{1}{2!}\langle T,T,D,D\rangle+ \frac{1}{2!}\langle T,T,H[D,D]\rangle + \langle T,D,H[T,D]\rangle + \ldots &= 0 \label{dilatonEqOld}
\end{aligned}
\ee
\be
\begin{aligned}
    &\frac{1}{3!} \langle D, D, D, D \rangle + \frac{1}{2!} \langle D,D,H[D,D]\rangle + \\
    &\frac{1}{3!} \langle D,T,T,T\rangle + \frac{1}{2!} \langle D,T,H[T,T]\rangle + \\
    &\frac{1}{2!} \langle D,D,T,T\rangle + \langle D,T,H[T,D]\rangle + \frac{1}{2!} \langle D,D,H[T,T]\rangle + \ldots &= 0 \end{aligned}
\ee
where we also made use of the symmetry of the closed string products.
Note that we can throw away the purely dilaton terms $\frac{1}{3!} \langle D, D, D, D \rangle + \frac{1}{2!} \langle D,D,H[D,D]\rangle $ since purely dilaton potential is flat \cite{Yang:2005ep} (this can be verified analytically to quartic order by methods similar to subsection \ref{elementary}). Once we introduce the four-point amplitudes $t^4 \mathcal{A}_{TTTT} \equiv \langle T,T,T,T \rangle + 3 \langle T,T,H[T,T]\rangle$, $t^2 d^2 \mathcal{A}_{TTDD} \equiv \langle T,T,D,D \rangle + \langle D,D,H[T,T]\rangle + 2 \langle T, D, H[T,D]\rangle$ and $t^3 d \, \mathcal{A}_{TTTD} \equiv \langle T,T,T,D \rangle + \langle D,T,H[T,T]\rangle + 2 \langle T,T,H[D,T]\rangle$ these equations simplify to
\be
\begin{aligned}
    &\langle T, Q T \rangle + \frac{1}{2!}\langle T,T,T\rangle + \frac{1}{3!} t^4 \mathcal{A}_{TTTT} + \frac{1}{2!} t^3 d \, \mathcal{A}_{TTTD} + \frac{1}{2!} t^2 d^2 \mathcal{A}_{TTDD} + \ldots &= 0 \label{tachyonEq}
\end{aligned}
\ee
\be
\begin{aligned}
    \frac{1}{3!} t^3 \mathcal{A}_{TTTD}  + \frac{1}{2!} t^2 d \, \mathcal{A}_{TTDD} + \ldots &= 0 \label{dilatonEq}
\end{aligned}
\ee
where we suppressed the purely dilaton part and used $\langle D,T,H[T,T]\rangle = \langle T,T,H[T,D]\rangle$ and in writing the dilaton equation of motion (\ref{dilatonEq}) we were a bit more careful by not writing a factor of $d$ since we actually overlap by $\frac{1}{d} D$ and not $D$ (otherwise $d=0$ would be a solution for nontrivial $t$, which it is not since interactions with the tachyons generate a tadpole for the dilaton). We note the importance of the $\frac{1}{3!}$ in (\ref{dilatonEq}), if it had been $\frac{1}{2!}$, the structure of the tachyon vacuum would not be changed by the dilaton.

Next we would like to compute the on-shell value of the action (\ref{action}) and to do this, we note that it simplifies thanks to the equations of motion
\be
S = \frac{1}{6}\biggr(\langle R, Q R\rangle - \frac{1}{2} \biggr[\frac{1}{2!}\langle R,R,H[R,R]\rangle + \frac{1}{3!} \langle R,R,R,R\rangle \biggr]\biggr) + \ldots
\ee
which means that we can evaluate the on-shell action to next-to-leading order with four-point amplitudes as
\be
S = \frac{1}{6}\biggr(\langle T,Q T\rangle - \frac{1}{12}(t^4 \mathcal{A}_{TTTT} + t^2 d^2 \mathcal{A}_{TTDD} + t^3 d \, \mathcal{A}_{TTTD})\biggr)+ \ldots
\label{actionOS}
\ee
keeping in mind that the fixed point couplings have to be plugged in. We now continue by evaluating the amplitudes $\mathcal{A}_{TTDD}$ and $\mathcal{A}_{TTTD}$ to leading order in $y$.

\subsection{Feynman contributions to dilaton amplitudes}
We start with some notation for the local coordinates on the-punctured sphere $\Sigma_{0,3}$. We write the local coordinates around $z_I$ as
\be
z = z_I + \rho w + \rho^2 \beta w^2 + \ldots
\ee
with $z_1 = 0$, $z_2= \infty$, $z_3 = 1$ the positions of the punctures, $\rho = \frac{1}{\lambda h'_3(1)}$ the mapping radius and we specialised to a symmetric vertex.

The quadratic and cubic couplings of the dilatons are zero on the level matched states except for the case of two dilatons and the identity since these vertices are only nonvanishing when the sum of the left and right ghost numbers of the insertions are equal. This means that $\langle T, D, H[T,D]\rangle = 0$ and $\langle T,T,H[D,T] \rangle=0$ so that $t^2 d^2 \mathcal{A}_{TTDD} = \langle T,T,D,D \rangle + \langle D,D,H[T,T]\rangle$ and $t^3 d\, \mathcal{A}_{TTTD} = \langle T,T,T,D \rangle$.

We now calculate $\langle D,D,H[T,T] \rangle$ by inserting the decomposition of the identity and keeping only terms nonvanishing in the $\lambda \to \infty$ limit, which gives
\be
\langle D,D,H[T,T]\rangle = \langle D,D, c_1\bar{c}_1 \rangle \langle c_1\bar{c}_1,T,T\rangle = \abs{\lambda h'_3(1)}^{2y+2} t^2 \langle D,D,c_1\bar{c}_1\rangle \,.
\ee
Note the extra minus from the propagator.
The difficulty now comes from the nonprimarity of the ghost dilaton $D = d (\frac{1}{2}c\del^2 c - h.c.)$
\be
f \circ (\frac{1}{2}c \del^2 c) - h.c. =\frac{1}{2} (f \circ c) \del^2 (f \circ c) = -\frac{f''}{2(f')^2} c \del c + \frac{1}{2} c \del^2 c - h.c.
\ee
which means that the local coordinates are needed to higher than leading order thanks to the presence of $f''$. Concretely for $f$ being a local coordinate map associated to a symmetric vertex
\be
f \circ (\frac{1}{2}c \del^2 c) - h.c. = - \beta c \del c + \frac{1}{2} c \del^2 c - h.c.
\ee
One then has
\be
\langle D,D,c_1\bar{c}_1\rangle = 2 \, d^2 \abs{\lambda h'_3(1)}^2 \langle -\beta c\del c + \frac{1}{2}c\del^2 c, c\rangle \langle\bar{\beta} \bar{c}\bar{\del}\bar{c} - \frac{1}{2} \bar{c}\bar{\del}^2\bar{c},\bar{c}\rangle = 2 \,d^2 \abs{\beta}^2 \abs{\lambda h'_3(1)}^2
\ee
since we normalise $\langle  c\del c,c\rangle \langle \bar{c}\bar{\del}\bar{c}, \bar{c}\rangle = 1$ and $\langle c\del^2 c,c\rangle = 0$.
This gives
\be
\langle D,D,H[T,T]\rangle = -2 \, d^2 t^2 \abs{\lambda h'_3(1)}^{2y + 4} \abs{\beta}^2 = -2 \, d^2 t^2 \abs{\lambda h'_3(1)}^4 \abs{\beta}^2 + O(y) \,.
\label{feynman}
\ee
Note that for polyhedral vertices $\abs{h'_3(1)} = \frac{3\sqrt{3}}{4}$ and $\abs{\beta} = \frac{1}{2}$ since when we compose the Witten vertex (see \cite{Erler:2019loq}) with an $SL(2,\cc)$ map $z\to \frac{z-e^{\frac{2\pi i}{3}}}{1-e^{\frac{2 \pi i}{3}} z}$, then for example the resulting local coordinate around 0 is $z = 0 - \frac{4}{3\sqrt{3}} w - \frac{8}{27} w^2 + \ldots$.

\subsection{Elementary contributions to dilaton amplitudes}
\label{elementary}
In this subsection, we calculate the elementary couplings $\langle T,T,D,D \rangle$ and $\langle T,T,T,D \rangle$. To do this, we need the local coordinates on the four-punctured sphere $\Sigma_{0,4}$. Around $z_I$ we write them as
\be
z = z_I(\xi,\bar{\xi}) + \rho_I(\xi,\bar{\xi}) w + \rho_I^2 \beta_I(\xi,\bar{\xi}) w^2 + \ldots
\ee
with $z_1(\xi,\bar{\xi}) = 0$, $z_2(\xi,\bar{\xi}) = \infty$, $z_3(\xi,\bar{\xi}) = 1$ and $z_4(\xi,\bar{\xi}) = \xi$.

In \cite{Bergman:1994qq} Bergman and Zwiebach find an expression for the form that one needs to integrate in order to compute the amplitude with one ghost dilaton (see \ref{TTTDIntegrand} for a derivation).
\be
\omega_D =  d \xi \wedge d \bar{\xi} \, (\del_{\bar{\xi}} \beta_4 + \del_\xi \bar{\beta_4})
\ee
This means that
\be
\langle T,T,T,D\rangle = -\threept \frac{1}{2\pi i} t^3 d \oint\displaylimits_{\del \mathcal{R}} ( -d \xi \, \beta_4 + d\bar{\xi} \,\bar{\beta_4}) + O(y)\label{tttdIntegral}
\ee
where we used the complex version of the Stokes theorem
\be
\int\displaylimits_\Omega d \xi \wedge d \bar{\xi} \biggr(\del_{\xi} f(\xi,\bar{\xi}) + \del_{\bar{\xi}} g(\xi,\bar{\xi})\biggr) = \oint\displaylimits_{\del \Omega} \biggr(d \bar{\xi} \, f(\xi,\bar{\xi}) - d \xi \, g(\xi,\bar{\xi})\biggr) \,.
\label{Stokes}
\ee
For polyhedral vertices this should give
\be
\label{tttd}
\langle T,T,T,D\rangle = t^3 d \, \mathcal{A}_{TTTD} = -t^3 d \, \threept + O(y)
\ee
since to leading order one can treat the $T$ punctures as fixed and then the theorem of \cite{Bergman:1994qq} says that $\mathcal{A}_{TTTD} \rvert_{T \, \mathrm{fixed}} = d \langle T,T,T\rangle$ where the constant of proportionality is the Euler characteristic of $\Sigma_{0,3}$ being equal to 1. We now check (\ref{tttd}) analytically. Note that by using the relations (\ref{b4c}), (\ref{cseries}) and transformation properties (\ref{b41}), (\ref{b4inf}) we have
\bea
\oint\displaylimits_{\del \mathcal{R}} d \xi \, \beta_4 &=& -\oint_{\abs{\xi}=\abs{\lambda h'_3(1)}^{-2}} d \xi \, \biggr[3\beta_4 - \frac{1}{\xi}\biggr] =-\oint_{\abs{\xi}=\abs{\lambda h'_3(1)}^{-2}} d \xi \, \frac{1}{2\xi}= -\pi i
\eea
where we Laurent expanded and kept only nonvanishing terms and included a minus sign from the opposite direction of normals when going from the left hand side of $\del\mathcal{R} = -\del\mathcal{R}_s  -\del\mathcal{R}_t -\del\mathcal{R}_u $ to the right hand side. Together with the conjugate contribution, one has
\be
\langle T,T,T,D\rangle = t^3 d \, \mathcal{A}_{TTTD} = -t^3 d \,\threept + O(y)
\ee
as we expected.

In \cite{Yang:2005ep} Yang and Zwiebach show that (see \ref{TTDDDer} for a derivation)
\be
\langle T,T,D,D\rangle = -t^2 d^2 \Re \frac{1}{2\pi i}\oint\displaylimits_\mathcal{\del \mathcal{R}} \biggr([(\del_\xi \beta_3)\bar{\xi}\bar{\beta}_4 -\beta_3\del_\xi(\bar{\xi}\bar{\beta}_4)]\, d\xi + [(\del_{\bar{\xi}}\beta_3)\bar{\xi}\bar{\beta}_4-\beta_3\del_{\bar{\xi}}(\bar{\xi}\bar{\beta}_4)] \,d\bar{\xi}\biggr) + O(y)
\label{ttdecko}
\ee
and by the dilaton theorem it is expected \cite{Yang:2005ep} to cancel with the Feynman contribution (\ref{feynman}) since to leading order our tachyon is marginal, leading to $\mathcal{A}_{TTDD} = O(y)$. We would like to rewrite the expression (\ref{ttdecko}) as an integral around $\mathcal{R}_u$ near zero. To do this, we need the transformation properties (\ref{b41})-(\ref{dinf}). First we transform the $t$-channel contribution (around infinity)
\bea
\langle T,T,D,D\rangle_\infty &=& - t^2 d^2 \Re \frac{1}{2\pi i}\oint\displaylimits_\infty \biggr([(\del_\xi \beta_3)\bar{\xi}\bar{\beta}_4 -\beta_3\del_\xi(\bar{\xi}\bar{\beta}_4)] \,d\xi + [(\del_{\bar{\xi}}\beta_3)\bar{\xi}\bar{\beta}_4-\beta_3\del_{\bar{\xi}}(\bar{\xi}\bar{\beta}_4)] \,d\bar{\xi}\biggr) \nonumber \\
			      &=& \langle T,T,D,D\rangle_0 - t^2 d^2 \Re \frac{1}{2\pi i} \oint\displaylimits_0 \biggr([-\del_\xi \beta_3 + \del_\xi(\bar{\xi}\bar{\beta}_4)]\,d\xi + [-\del_{\bar{\xi}}\beta_3 + \del_{\bar{\xi}}(\bar{\xi}\bar{\beta}_4)]\,d\bar{\xi}\biggr) \nonumber
\eea
and continue with the $s$-channel (around 1)
\bea
\langle T,T,D,D\rangle_1 &=& - t^2 d^2 \Re \frac{1}{2\pi i}\oint\displaylimits_1 \biggr([(\del_\xi \beta_3)\bar{\xi}\bar{\beta}_4 -\beta_3\del_\xi(\bar{\xi}\bar{\beta}_4)] \, d\xi + [(\del_{\bar{\xi}}\beta_3)\bar{\xi}\bar{\beta}_4-\beta_3\del_{\bar{\xi}}(\bar{\xi}\bar{\beta}_4)] \,d\bar{\xi}\biggr) \nonumber \\
			 &=& - t^2 d^2 \Re \frac{1}{2\pi i} \oint\displaylimits_0 \biggr([(\del_\xi \beta_1)(1-\bar{\xi})\bar{\beta}_4 + (1-\beta_1)\del_\xi((1-\bar{\xi})\bar{\beta}_4)] \,d\xi + \\ && \hspace{2.3 cm}[(\del_{\bar{\xi}} \beta_1)(1-\bar{\xi})\bar{\beta}_4 + (1-\beta_1)\del_{\bar{\xi}}((1-\bar{\xi})\bar{\beta}_4)] \,d\bar{\xi}\biggr) \nonumber \,.
\eea
When one plugs in (\ref{b1c}), (\ref{b3c}) and (\ref{b4c}) together with the series expansion (\ref{cseries}), one gets
\bea
\langle T,T,D,D\rangle_0 &=& 0 \\
\langle T,T,T,T\rangle_\infty &=& 0 \\
\langle T,T,D,D\rangle_1 &=& \frac{1}{2} t^2 d^2\,  \abs{\lambda h'_3(1)}^4
\eea
which results in a cancellation with the Feynman contribution (\ref{feynman}) since $\abs{\beta_3}^2 = \frac{1}{4}$
\be
\mathcal{A}_{TTDD} = -2 \frac{1}{4} \abs{\lambda h'_3(1)}^4 +\frac{1}{2} \abs{\lambda h'_3(1)}^4 + O(y) = O(y)
\ee
as we expected.

\subsection{Analysing the solution}
Let us now summarise the effect of adding the ghost dilaton. First of all, it allowed us to avoid the obstruction. However, it came at a cost as it is impossible to solve the equations (\ref{tachyonEq}) and (\ref{dilatonEq}) with $d$ of order $O(y)$. This is important as it removes an obvious contradiction with the theorem by Erler \cite{Erler:2022agw} that the CSFT action should vanish on-shell.

It is clear from (\ref{actionOS}), that if one had $t = O(y)$ and $d = O(y)$, then the action would not vanish to order $O(y^3)$ so the only hope for the result of \cite{Erler:2022agw} to hold is for at least one of the couplings to become large. To see that it is indeed what happens, we simply look at (\ref{dilatonEq}) while using $\mathcal{A}_{TTTD} = - \threept + O(y)$, $\mathcal{A}_{TTDD} = O(y)$ and since the neglected terms are $O(y^4)$, we get a violation of the $O(y^3)$ part of the equation of motion from the $t^3$ term. We note that by the dilaton theorem any amplitude with two tachyons and two or more dilatons should be $O(y)$.

Thus by contradiction at least one of the couplings must be large, that is either $t$ or $d$ must be $O(1)$. If $t = O(1)$ with $d = O(y)$ would hold then the $O(1)$ parts of (\ref{tachyonEq}) and (\ref{dilatonEq}) give an overdetermined system for $t$, which we think is unlikely to have a solution.

We are left with the possibilities of $t = O(y)$ with $d = O(1)$ or $t = O(1)$ with $d = O(1)$. The latter possibility would require that the term explicitly proportional to $y$ in (\ref{tachyonEq}) is zero. This would hold only with a cancellation of the type $t^2 (\frac{y}{2} + \frac{1}{2!}d^2 \mathcal{A}_{TTDD} + \frac{1}{3!}d^3 \mathcal{A}_{TTDDD}+\ldots) = 0$, which means that there would need to be some universal $d$ that cancels the bracket (the amplitudes with two tachyons are not sensitive to $\threept$ etc.). But from (\ref{dilatonEq}) we see that $d$ should depend on matter so that we conclude $t = O(y)$ and $d = O(1)$.

Last, we remark that if we truncate the sum over the dilaton contributions to the value of the on-shell action (\ref{actionOS}), there is no reason for it to vanish. For that we would presumably need to sum over all possible dilaton insertions. It would be nice to excplicitly work out the dilaton couplings so the result of \cite{Erler:2022agw} can be explicitly confirmed.

\sectiono{Conclusions and outlook}
\label{conclusion}
In this paper we studied the condensation of nearly marginal matter tachyons in the framework of CSFT. Our main finding was the observation that upon ommiting the ghost dilaton from the spectrum, the on-shell value of the CSFT action computed the shift in the central charge $\Delta c$ between the initial background and the tachyon vacuum via $S = -\frac{1}{12} \Delta c$. This led to a novel conformal perturbation theory formula (\ref{cecko}) for $\Delta c$, which we tested on Zamolodchikov's flow between consecutive minimal models. Upon reintroduction of the ghost dilaton, the previously found tachyon vacuum seems to get finitely rescaled and one cannot simply conclude that the action is nonzero, avoiding immidiate contradiction with \cite{Erler:2022agw}.

This rather unexpected observation has left many questions open, below we list some of them.
\begin{enumerate}
	\item Can we find some physical mechanism that justifies truncating the CSFT spectrum ?
		\begin{itemize}
			\item This may be connected to the fact that the ghost dilaton changes the string coupling constant while keeping the underlying CFT unchanged \cite{Bergman:1994qq} and thus for the purposes of studying CFTs, truncation of the spectrum might be sensible.
		\end{itemize}
	\item Is the resulting theory consistent ?
		\begin{itemize}
			\item Specifically, we ask whether it inherits some of the gauge invariance of the parent theory. One could perhaps use some of the homotopy transfer technology \cite{Erbin:2020eyc}. However it would be difficult to physically interpret this theory since the resulting matter CFT with central charge $26 - \Delta c$ does not give a valid critical string background. It's unclear where the Liouville direction would come from if not included from the beginning \cite{Mukherji:1991tb}.
		\end{itemize}
	\item Can we extract some lessons about general (possibly nonperturbative) CSFT solutions or is the structure we found simply a peculiarity of our setup ?
	\begin{itemize}
		\item A naive application to bulk tachyon condensation would indicate that the tachyon potential without the ghost dilaton contributions should make sense, but this is in tension with its dubious convergence properties \cite{Yang:2005rx, Moeller:2006cv, Moeller:2006cw}. Perhaps it is relevant that if the tachyon vacuum has empty cohomology, then the string coupling becomes unobservable since there no longer is a ghost dilaton \cite{Belopolsky:1995vi}.
	\end{itemize}
	\item Is it possible to better understand the dilaton contributions ?
	\begin{itemize}
		\item In particular, it would be interesting to develop a sort of soft ghost dilaton theorem for nearly on-shell amplitudes, which would allow us to sum up all amplitudes with a fixed number of tachyons but all possible numbers of dilatons.
		\item In our setup it seems that the ghost dilaton is a major driving force behind moving the on-shell action close to zero. Is this part of the reason why the couplings to the ghost dilaton restore the minimum of the bulk tachyon potential \cite{Yang:2005rx, Moeller:2006cv} ?
	\end{itemize}
	\item From the point of view of conformal perturbation theory, SFT with large stubs provided us with a very efficient SFT-inspired point-splitting regularisation scheme at quartic order, can it be pushed to higher orders ?
	\begin{itemize}
		\item It seems that this could be particularly fruitful in the context of boundary conformal perturbation theory, where the quartic order calculation is shockingly simple, see appendix \ref{osft} and compare it to \cite{Scheinpflug:2023osi}.
		\item One could also use SFT to find formulae for quantities of interest in conformal perturbation theory other than the shift $\Delta c$. An important next step might be to compute the shift in the spectrum, see \cite{Sen:2014dqa, Sen:2019jpm, Maccaferri:2022yzy}.
	\end{itemize}
\end{enumerate}

\section*{Acknowledgments}

We are grateful to Georg Stettinger for collaboration at the early stages of this work and to Atakan Hilmi-Firat for explaining several aspects of string vertices to us. We also thank Ted Erler, Jakub Vošmera and Barton Zwiebach for discussions on the physics of the ghost dilaton and Tomáš Procházka for pointing us towards the relevant conformal perturbation theory literature. This work
has been supported by Grant Agency of the Czech Republic, under the grant EXPRO 20-25775X

\appendix
\sectiono{Regularising the four-tachyon interaction}
\label{reg}
In this appendix, we derive the formula (\ref{fourtachyon}) for the finite part $\mathcal{A}_{TTTT}^{f.p.}$ of the four-tachyon amplitude $\mathcal{A}_{TTTT}$ with $T = t c_1\bar{c}_1 V\ket{0}$, where $V$ is an $(h,h)$ matter primary with $y = 2(1-h)$ small. It is defined by
\be
\frac{1}{t^4}\langle T,T,T,T\rangle = 3 \abs{\lambda h'_3(1)}^{4} + 12 \threept^2 \ln\abs{\lambda h'_3(1)} + \mathcal{A}_{TTTT}^{f.p.},
\label{div}
\ee
see (\ref{amp}). The elementary coupling $\langle T,T,T,T\rangle$ is given by
\be
\frac{1}{t^4} \langle T,T,T,T\rangle = -\frac{1}{2\pi i} \int\displaylimits_\mathcal{R} d\xi \wedge d\bar{\xi} \, \, \bra{V} V(1) V(\xi,\bar{\xi}) \ket{V},
\ee
with $\mathcal{R}$ being the three-punctured sphere with excisions around punctures, see (\ref{choicev}).
It should be possible to extend the integration from the vertex region $\mathcal{R}$ to the entire complex plane if we somehow split the integral to two parts, where one would be manifestly finite and the other would reproduce the divergence structure of (\ref{div}). To do this, we expand into conformal blocks and expand those around zero \cite{DiFrancesco:1997nk} (note the fusion $V \cross V = 1 + V$)
\bea
\bra{V} V(1) V(\xi,\bar{\xi}) \ket{V} = \frac{1}{\abs{\xi}^4}\biggr(1+ \frac{2}{c} \xi^2 + \ldots\biggr)\biggr(1+\frac{2}{c} \overline{\xi}^2+\ldots\biggr) + \threept^2 \frac{1}{\abs{\xi}^2} (1 + \ldots)(1 + \ldots) + O(y) \nonumber
\eea
where correctly we should have written $\threept^2 \frac{1}{\abs{\xi}^2} (1 + \frac{1}{2} \xi \ldots)(1 + \frac{1}{2} \bar{\xi}\ldots)$ but this gives the same result since the difference is a vanishing lens contribution, see (\ref{lens}).
Now we could naively subtract for example $\frac{1}{\abs{\xi}^4}$ and $\frac{1}{\abs{\xi-1}^4}$ to cure the leading singularity around zero and one. But this would give ill-defined behavior at infinity since both of these terms are regular there and we know by permutation symmetry that a subtraction at infinity is also needed. The solution is to subtract $\frac{1}{\abs{\xi}^4}\frac{1}{\abs{\xi-1}^4}$ and then permute the punctures by $Sl(2,\cc)$ maps (note that one has to add a factor of $\frac{1}{2}$ to avoid overcounting subtractions). \\ Proceeding analogously for the other singularities, we get
\bea
\frac{1}{t^4} \langle T,T,T,T\rangle &=& -\frac{1}{2\pi i} \biggr(\int\displaylimits_\cc d\xi \wedge d\bar{\xi} \, \, \biggr[ \bra{V} V(1) V(\xi,\bar{\xi}) \ket{V} - \frac{1}{2}\biggr(\frac{1}{\abs{\xi}^4\abs{\xi-1}^4} + \frac{\abs{\xi-1}^4}{\abs{\xi}^4} + \frac{\abs{\xi}^4}{\abs{\xi-1}^4} \nonumber \\\ && \hspace{1 cm}-4 \biggr[\frac{1}{\abs{\xi}^2\abs{\xi-1}^2} + \frac{1}{\abs{\xi}^2} + \frac{1}{\abs{\xi-1}^2}\biggr]\biggr)  - \frac{2}{c} \biggr(\frac{1}{(\xi-1)^2} + \frac{1}{(\bar{\xi}-1)^2}\biggr) \nonumber \\ && \hspace{1 cm} - \frac{\threept^2}{2}\biggr( \frac{1}{\abs{\xi}^2\abs{\xi-1}^2} + \frac{1}{\abs{\xi}^2} + \frac{1}{\abs{\xi-1}^2}\biggr) \biggr] \nonumber \\ && \hspace{1cm} + \int\displaylimits_\mathcal{R} d\xi \wedge d\bar{\xi} \, \, \biggr[ \frac{1}{2}\biggr(\frac{1}{\abs{\xi}^4\abs{\xi-1}^4} + \frac{\abs{\xi-1}^4}{\abs{\xi}^4} + \frac{\abs{\xi}^4}{\abs{\xi-1}^4} \nonumber \\\ && \hspace{1 cm}-4 \biggr[\frac{1}{\abs{\xi}^2\abs{\xi-1}^2} + \frac{1}{\abs{\xi}^2} + \frac{1}{\abs{\xi-1}^2}\biggr]\biggr)  + \frac{2}{c} \biggr(\frac{1}{(\xi-1)^2} + \frac{1}{(\bar{\xi}-1)^2}\biggr) \nonumber \\ && \hspace{1 cm} + \frac{\threept^2}{2}\biggr( \frac{1}{\abs{\xi}^2\abs{\xi-1}^2} + \frac{1}{\abs{\xi}^2} + \frac{1}{\abs{\xi-1}^2}\biggr) \biggr]\biggr) + O(y)
\label{tecka}
\eea
where we subtracted and added the same contribution while for the finite contribution, we changed the integration region from $\mathcal{R}$ to $\cc$. We also had to subtract $4 \biggr[\frac{1}{\abs{\xi}^2\abs{\xi-1}^2} + \frac{1}{\abs{\xi}^2} + \frac{1}{\abs{\xi-1}^2}\biggr]$ since the subtraction $\frac{1}{\abs{\xi}^4\abs{\xi-1}^4} + \frac{\abs{\xi-1}^4}{\abs{\xi}^4} + \frac{\abs{\xi}^4}{\abs{\xi-1}^4}$ actually generates a subleading divergence itself. Note that subtractions of similar form as those in (\ref{tecka}) already appeared in \cite{Komargodski:2016auf}. It also turns out that in our analysis the analogue of the subtraction
\be
\frac{2}{c} \frac{1}{\abs{\xi}^4} (\xi^2+\overline{\xi}^2) = \frac{2}{c} \biggr(\frac{1}{\xi^2} + \frac{1}{\overline{\xi}^2}\biggr)
\ee
only has to be subtracted around one since there certain lens-like shapes of (\ref{lens}) live (when integrated on a region with angular symmetry, it does not give a nontrivial contribution but the lenses do not have such symmetry).
\begin{figure}[h!]
	\centering
	\includegraphics[width=7cm]{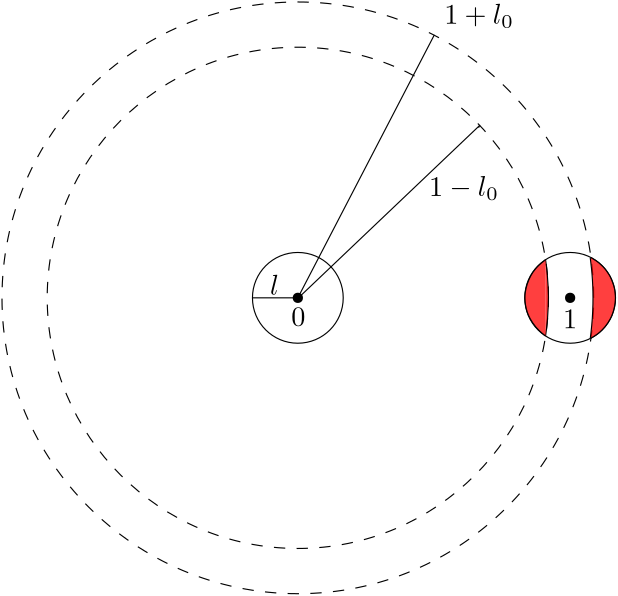}
	\caption{The lens-like regions are depicted in red}
	\label{lensfig}
\end{figure}

Now we explicitly carry out the residual integration over $\mathcal{R}$. We would like to integrate radially outwards from zero, but the disc around one is standing in our way. To remedy this, we pretend that this disc does not exist and we then account for it separately by subtracting integrals over certain lens-like subregions, see figure \ref{lensfig} and appendix \ref{lens}. In this figure, we introduce a small (parametrically smaller than the size of our Feynman regions) auxiliary parameter $l_0$, which prevents us from integrating straight through one. We also write $l \equiv \frac{1}{\abs{\lambda h'_3(1)}^2}$ and note that the parameter $l_0$ must cancel out of our results since it is only a relict of a convenient choice of integration coordinates. We expect that the integration over $\mathcal{R}$ produces the divergences present in (\ref{div}) and some finite part, which contributes to $\mathcal{A}_{TTTT}^{f.p.}$. Lastly, note that $d\xi \wedge d\bar{\xi} = -2i r dr \wedge d\varphi$ so that the resulting measure with respect to which we integrate the subtractions is
		\be
		-\frac{1}{2\pi i} (-2i) r dr\wedge d\varphi = \frac{1}{\pi} r dr\wedge d\varphi
		\ee

\begin{enumerate}
	\item Contribution of  $I_1 \equiv \frac{1}{2}\biggr(\frac{1}{\abs{\xi}^4\abs{\xi-1}^4} + \frac{\abs{\xi-1}^4}{\abs{\xi}^4} + \frac{\abs{\xi}^4}{\abs{\xi-1}^4}  -4 \biggr[\frac{1}{\abs{\xi}^2\abs{\xi-1}^2} + \frac{1}{\abs{\xi}^2} + \frac{1}{\abs{\xi-1}^2}\biggr]\biggr)$

Upon performing the angular integration, one obtains
\bea
\frac{1}{\pi}\int\displaylimits_0^{2\pi} d\varphi \, r I_1 &=& -\frac{2  \left(4 r^4-3 r^2+1\right)}{r^3 \left(r^2-1\right)^3}  \, \, \, \, r < 1 \nonumber \\
							   &=&  \frac{2  r^3 \left(r^4-3 r^2+4\right)}{\left(r^2-1\right)^3}\, \, \, \, \hspace{0.15 cm}r > 1 \nonumber
\eea
Now integrating over two annuli as in figure \ref{lensfig} gives
\bea
\int\displaylimits_{l}^{1-l_0} -\frac{2 \left(4 r^4-3 r^2+1\right)}{r^3 \left(r^2-1\right)^3} d r + \int\displaylimits_{1-l_0}^{\frac{1}{l}} \frac{2 r^3 \left(r^4-3 r^2+4\right)}{\left(r^2-1\right)^3} d r \sim \frac{2 }{l^2}+\frac{1 }{2 l_0^2}-\frac{17 }{8}
\eea
where we performed the limit $l \sim 0$ with $\frac{l_0}{l} \sim 0$. Since $I_1$ behaves as $\frac{1}{\abs{\xi-1}^4}$ around 1, the lens contribution we need to subtract is (\ref{sol1}) (divided by $\pi$) resulting in the total contribution
\be
\frac{2 }{l^2}+\frac{1}{2 l_0^2}-\frac{17 }{8} - \biggr(-\frac{1}{l^2} + \frac{1}{2 l_0^2} - \frac{1}{8} \biggr) = \frac{3}{l^2} - 2
\ee
The divergent part is exactly the same as required by (\ref{div}) and the $-2$ contributes to $\mathcal{A}_{TTTT}^{f.p.}$.

\item Contribution of  $I_2 \equiv \frac{2}{c} \biggr(\frac{1}{(\xi-1)^2} + \frac{1}{(\bar{\xi}-1)^2}\biggr) $

	Upon performing the angular integration, one obtains
	\bea
\frac{1}{\pi}\int\displaylimits_0^{2\pi} d\varphi \, r I_2 &=& \frac{2}{c} 4 r \, \, \, \, r < 1 \nonumber \\
							   &=&  0\, \, \, \, \, \, \hspace{0.35 cm} r > 1 \nonumber
\eea
so that one has a contribution only from one annulus
\be
\int\displaylimits_{l}^{1-l_0} \frac{2}{c} 4 r d r\sim \frac{2}{c} 2
\ee
and subtracting (\ref{sol3}), we get
\be
\frac{2}{c} (2-2) = 0
\ee
meaning that $I_2$ does not contribute when integrated over $\mathcal{R}$. The contribution of $-I_2$ from the subtraction can be integrated over $\cc$, giving $-\frac{4}{c}$.

\item Contribution from $I_3\equiv \frac{\threept^2}{2}\biggr( \frac{1}{\abs{\xi}^2\abs{\xi-1}^2} + \frac{1}{\abs{\xi}^2} + \frac{1}{\abs{\xi-1}^2}\biggr)$

	Upon performing the angular integration, one obtains
\bea
\frac{1}{\pi}\int\displaylimits_0^{2\pi} d\varphi \, r I_3 &=& -\threept^2 \frac{2  }{r \left(r^2-1\right)}  \, \, \, \, r < 1 \nonumber \\
							   &=&  \threept^2 \frac{2  r}{r^2-1}\, \, \, \, \hspace{0.85 cm}r > 1 \nonumber
\eea
Inegrating over both annuli, we get
\bea
\int\displaylimits_{l}^{1-l_0} -\threept^2 \frac{2  }{r \left(r^2-1\right)} d r + \int\displaylimits_{1-l_0}^{\frac{1}{l}} \threept^2  \frac{2  r}{r^2-1} d r \sim \threept^2(-2 \ln 2 -4 \ln l -2 \ln l_0)
\eea
and subtracting the contribution from the lenses (\ref{sol2}) gives
\be
\threept^2 \biggr(-2 \ln 2 -4 \ln l -2 \ln l_0 - 2 \ln \frac{l}{2 l_0} \biggr)= -6 \threept^2 \ln l
\ee
which covers the logarithmic divergence in (\ref{div}) and does not produce a finite part.
\end{enumerate}

Taking all these three contributions together reproduces the expected divergences and results in only $-2$ and $-\frac{4}{c}$ being added to the finite part and thus
\bea
\mathcal{A}_{TTTT}^{f.p.} &=& -\frac{1}{2\pi i} \int\displaylimits_\cc d\xi \wedge d\bar{\xi} \, \, \biggr[ \bra{V} V(1) V(\xi,\bar{\xi}) \ket{V} - \frac{1}{2}\biggr(\frac{1}{\abs{\xi}^4\abs{\xi-1}^4} + \frac{\abs{\xi-1}^4}{\abs{\xi}^4} + \frac{\abs{\xi}^4}{\abs{\xi-1}^4} \nonumber \\\ && \hspace{1 cm}-4 \biggr[\frac{1}{\abs{\xi}^2\abs{\xi-1}^2} + \frac{1}{\abs{\xi}^2} + \frac{1}{\abs{\xi-1}^2}\biggr]\biggr)  - \frac{\threept^2}{2}\biggr( \frac{1}{\abs{\xi}^2\abs{\xi-1}^2} + \frac{1}{\abs{\xi}^2} + \frac{1}{\abs{\xi-1}^2}\biggr) \biggr] \nonumber \\ &&  - 2 - \frac{4}{c}
\eea
concluding the derivation.

\sectiono{Integration over lens-like regions}
\label{lens}
In this appendix we compute integrals over certain lens-like regions, essentially going over the analysis of \cite{Poghossian:2013fda}. When computing the vertex region contribution $\langle T,T,T,T\rangle$ to the four tachyon amplitude $\mathcal{A}_{TTTT}$, we encounter that upon going to radial coordinates one accidentaly integrates over lens-like subregions $\mathcal{D}_L$ and $\mathcal{D}_R$ of the Feynman region around 1 (these should not be present since the Feynman and vertex regions are separate). We thus need to subtract the contributions from these subregions. For an illustration see figure \ref{lensfig}.

\begin{figure}[h!]
	\centering
	\includegraphics[width=7cm]{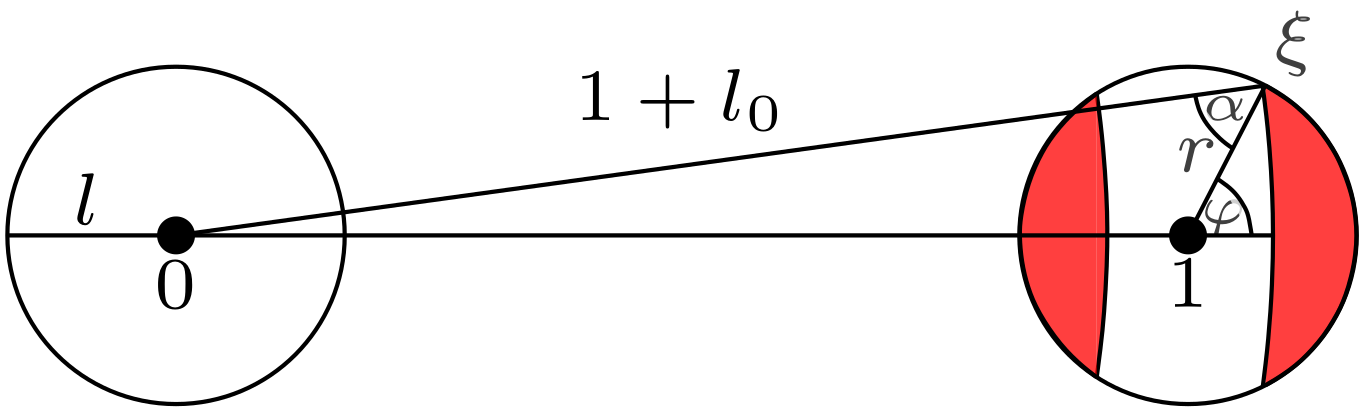}
	\caption{The lens-like regions are depicted in red}
	\label{lensfigangle}
\end{figure}

Concretely what we'll need to subtract are
\bea
			 &&\int\displaylimits_{\mathcal{D}_L \cup \dd_R} d\xi \wedge d\bar{\xi} \, \frac{1}{\abs{\xi-1}^4} \label{int1}
		      \\ && \int\displaylimits_{\mathcal{D}_L \cup \dd_R} d\xi \wedge d\bar{\xi} \, \frac{1}{\abs{\xi-1}^2} \label{int2}
		      \\ && \int\displaylimits_{\mathcal{D}_L \cup \dd_R} d\xi \wedge d\bar{\xi} \, \biggr(\frac{1}{(\xi-1)^2} + \frac{1}{(\bar{\xi}-1)^2}\biggr) \label{int3}
\eea
To do this, we restrict ourselves to $\mathcal{D}_R$ and all formulas valid for it will be valid for $\mathcal{D}_L$ upon $l_0 \to -l_0$. We'll need some geometry associated with the triangles in (\ref{lensfigangle}).

\bea
(1 + l_0)\sin \alpha &=& \sin \varphi \label{geo1} \\
r &=& (1+l_0)\cos \alpha - \cos \varphi \label{geo2}
\eea
where the first equality follows from the sine theorem and the second one follows from continuing the side of length $r$ while keeping the side of length $1+l_0$ fixed until a right angle is made.
From (\ref{geo1}) and (\ref{geo2}) it follows that
\be
\frac{1}{r} = \frac{(1+l_0)\cos \alpha + \cos \varphi}{l_0(l_0+2)}
\ee
We now use Stokes theorem to simplify the integrals (\ref{int1})-(\ref{int3})
\bea
\int\displaylimits_{\dd_R} d\xi \wedge d\bar{\xi} \, \frac{1}{\abs{\xi-1}^4} &=& - \frac{1}{2} \oint\displaylimits_{\del\mathcal{D}_R} d \varphi \, \frac{1}{r^2} \\
\int\displaylimits_{\dd_R} d\xi \wedge d\bar{\xi} \, \frac{1}{\abs{\xi-1}^2} &=& \oint\displaylimits_{\del\mathcal{D}_R} d \varphi \, \ln r  \\
\int\displaylimits_{\dd_R} d\xi \wedge d\bar{\xi} \, \biggr(\frac{1}{(\xi-1)^2} + \frac{1}{(\bar{\xi}-1)^2}\biggr) &=& - \oint\displaylimits_{\del\mathcal{D}_R} d r \, \frac{\sin(2\varphi)}{r}
\eea
Now we evaluate these integrals in the limit of $l \sim 0$ with $\frac{l_0}{l} \sim 0$ so that $\alpha \sim \varphi \sim \pm \frac{\pi}{2}$ on the top and bottom intersections respectively.
The boundary $\del \mathcal{D}_R$ splits into a part $\del \mathcal{D}_R^r$ where $\abs{\xi - 1} = r$ and $\del \mathcal{D}_R^l$ where $\abs{\xi-1} = l$ with the integration over the latter being trivial since it is a circular arc
\bea
\int\displaylimits_{\dd_R^l} d\xi \wedge d\bar{\xi} \, \frac{1}{\abs{\xi-1}^4} &\sim& -\frac{\pi}{2l^2} \\
\int\displaylimits_{\dd_R^l} d\xi \wedge d\bar{\xi} \, \frac{1}{\abs{\xi-1}^2} &\sim& \pi \ln l \label{analog} \\
\int\displaylimits_{\dd_R^l} d\xi \wedge d\bar{\xi} \, \biggr(\frac{1}{(\xi-1)^2} + \frac{1}{(\bar{\xi}-1)^2}\biggr) &\sim& 0
\eea

We observe that on the other part of the boundary $\del \mathcal{D}^r$ we can use Stokes theorem again on two of the three integrals since
\bea
d\varphi \frac{1}{r^2} &=& d\varphi \frac{(1+l_0)^2\cos^2\alpha+2(1+l_0)\cos\alpha \cos \varphi + \cos^2\varphi}{l_0^2(l_0+2)^2} \nonumber \\ &=& \frac{(1+l_0)^2d(\alpha+\varphi)+(1+l_0) d\sin(\alpha+\varphi)}{l_0^2(l_0+2)^2} \\
d r \frac{\sin(2\varphi)}{r} &=& \frac{(1+l_0)^2}{2} d(2\alpha - \sin(2\alpha))
\eea
Again expanding around $l \sim 0$, $l_0 \sim 0$ with $\frac{l_0}{l} \sim 0$ we get
\bea
\int\displaylimits_{\dd_R^r} d\xi \wedge d\bar{\xi} \, \frac{1}{\abs{\xi-1}^4} &\sim& \frac{\pi}{4 l_0^2} + \frac{\pi}{4 l_0} - \frac{\pi}{16} \\
\int\displaylimits_{\dd_R^r} d\xi \wedge d\bar{\xi} \, \biggr(\frac{1}{(\xi-1)^2} + \frac{1}{(\bar{\xi}-1)^2}\biggr) &\sim& \pi
\eea
where we note that one has to be careful about orientation and that the integrals have been done in terms of elementary functions but we opted out of writing the full expressions. The integral $\int\displaylimits_{\del\mathcal{D}^r_R} d \varphi \, \ln r$ cannot be expressed in terms of elementary functions but one can show
\be
\int\displaylimits_{\dd_R^r} d\xi \wedge d\bar{\xi} \, \frac{1}{\abs{\xi-1}^2} \sim -\pi \ln (2 l_0) \\
\ee
where the minus comes from opposite orientation. Accouting for the left lens and putting everything together then gives

\bea
\int\displaylimits_{\mathcal{D}_L \cup \dd_R} d\xi \wedge d\bar{\xi} \, \frac{1}{\abs{\xi-1}^4} &\sim& -\frac{\pi}{l^2} - \frac{\pi}{8} + \frac{\pi}{2 l_0^2} \label{sol1}
											    \\  \int\displaylimits_{\mathcal{D}_L \cup \dd_R} d\xi \wedge d\bar{\xi} \, \frac{1}{\abs{\xi-1}^2} &\sim& 2\pi \ln \frac{l}{2 \l_0}\label{sol2}
										    \\  \int\displaylimits_{\mathcal{D}_L \cup \dd_R} d\xi \wedge d\bar{\xi} \, \biggr(\frac{1}{(\xi-1)^2} + \frac{1}{(\bar{\xi}-1)^2}\biggr) &\sim&  2\pi \label{sol3}
\eea

\sectiono{Extracting local coordinates from quadratic differentials}
\label{diff}
Below we briefly review the procedure of extracting local coordinates from quadratic differentials for the case of the four punctured sphere \cite{Moeller:2004yy}. We write the quadratic differential as
\be
\varphi = \sum_{i=1}^4\biggr[\frac{-1}{(z-\xi_i)^2}+ \frac{c_i}{z-\xi_i}\biggr] dz^2
\ee
where $c_i$ is called the \textit{accessory parameter}. Let's say that we know the accessory parameter $c \equiv c_4$ for the moving puncture, then by the condition of regularity at $z = \infty$, we have
\be
\varphi = \biggr[\frac{-z^4+a z^3 + (2\xi -(1+\xi)a)z^2+a \xi z-\xi^2}{z^2(z-1)^2(z-\xi)^2}\biggr] dz^2
\ee
with $a = 2 + \xi(\xi-1) c$.
The local coordinates can be obtained by expanding the quadratic differential around the punctures so that upon expansion around $\xi_n$
\be
\varphi = \biggr[-\frac{1}{(z-\xi_n)^2} + \frac{2 \beta_n}{z-\xi_n} + \ldots\biggr] dz^2\,.
\ee
This gives
\bea
\beta_1(\xi,\bar{\xi}) &=& \frac{a}{2\xi}-\frac{1}{\xi}-1 =\frac{1}{2} (\xi-1)c-1 \label{b1c} \\
\beta_2(\xi,\bar{\xi}) &=& \frac{a}{2}-1-\xi = \frac{1}{2}\xi(\xi-1) c \\
\beta_3(\xi,\bar{\xi}) &=& \frac{a(\xi,\bar{\xi})-2 \xi}{2(1-\xi)}= 1 -\frac{1}{2}\xi c(\xi,\bar{\xi}) \label {b3c} \\
\beta_4(\xi,\bar{\xi}) &=& \frac{2-a(\xi,\bar{\xi})}{2(1-\xi)\xi} = \frac{1}{2}c(\xi,\bar{\xi}) \label{b4c}
\eea
with $c$ being left undetermined.

We are interested in quadratic differentials that are Jenkins-Strebel differentials describing local coordinates inside the Feynman regions. Luckily for us, the accessory parameter for this case has been bootstrapped in \cite{Firat:2023glo} as an expansion around the degeneration region $\xi = 0$ (the Strebel differential describing the vertex region has been bootstrapped as well but we won't need it). Concretely one has
\be
\label{cseries}
c(\xi,\bar{\xi}) = -4 \del_\xi f_1 (\xi) = \frac{1}{\xi} -\frac{1}{2} - \frac{37}{32} \xi - \frac{95}{64}\xi^2 + \ldots
\ee
with $f_1(\xi) = -\frac{\log (\xi )}{4} +\frac{\xi }{8} + \frac{37 \xi ^2}{256} +  \frac{95 \xi ^3}{768}+\ldots$ being a classical conformal block.
Note that one can obtain the corresponding expansions around $1$ and $\infty$ by using $a(1-\xi) = 4- a(\xi)$ and $a(\frac{1}{\xi}) = \frac{a(\xi)}{\xi}$. These give \cite{Yang:2005ep}
\bea
\beta_4(1-\xi) &=& -\beta_4(\xi) \label{b41} \\
\beta_4\biggr(\frac{1}{\xi}\biggr) &=& \xi[1-\xi \beta_4(\xi)] \label{b4inf}
\eea
\bea
\beta_3(1-\xi) &=& 1 - \beta_1(\xi) = 2 + \frac{1}{2}(1-\xi)c \\ \label{b31}
\beta_3\biggr(\frac{1}{\xi}\biggr) &=& 1-\beta_3(\xi) = \frac{1}{2}\xi c \label{b3inf} \,.
\eea
Also note the transformation laws for derivatives
\bea
\del_{1-\xi} &=& - \del_\xi \\ \label{d1}
\del_{\frac{1}{\xi}} &=& -\xi^2 \del_\xi \label{dinf} \,.
\eea

\sectiono{Vertex region integrands for interactions with dilatons}
\label{integrands}
The vertex region contribution to a general off-shell four-point amplitude is \cite{Yang:2005ep}
\be
\langle \Phi_1, \Phi_2,\Phi_3,\Phi_4\rangle \equiv -\frac{1}{2\pi i}\int \displaylimits_\mathcal{R} d \xi \wedge d \bar{\xi} \bra{\Sigma_{0,4}} \mathcal{B} \mathcal{B}^\star \ket{\Phi_1}\ket{\Phi_2}\ket{\Phi_3}\ket{\Phi_4}
\ee
with $\bra{\Sigma_{0,4}}$ being a surface state corresponding to $\Sigma_{0,4}$ so that $\bra{\Sigma_{0,4}}\ket{\Phi_1}\ket{\Phi_2}\ket{\Phi_3}\ket{\Phi_4} = \langle \Phi_1\Phi_2\Phi_3\Phi_4\rangle_{\Sigma_{0,4}}$ and the $b$-ghost insertions are
\be
\mathcal{B} = \sum_{I=1}^4\sum_{m=-1}^\infty (B_m^I b_m^I + \overline{C}_m^I\bar{b}_m^I), \, \, \,
\mathcal{B}^\star = \sum_{I=1}^4\sum_{m=-1}^\infty (\overline{B}_m^I \bar{b}_m^I + C_m^I b_m^I)
\ee
where
\be
B_m^I = \oint \frac{d w}{2\pi i} \frac{1}{w^{m+2}} \del_\xi h_I [\del_w h_I]^{-1}, \, \, \, \, C_m^I = \oint \frac{d w}{2\pi i} \frac{1}{w^{m+2}} \del_{\bar{\xi}} h_I [\del_w h_I]^{-1}.
\ee
and the $\star$-conjugation is a complex conjugation on numbers and turns holomorphic ghosts into antiholomorphic ghosts while reversing their order when one considers a product a ghosts.
The local coordinates around the $I$-th puncture are expanded as
\be
z = h_I(w,\xi,\bar{\xi}) = z_I(\xi,\bar{\xi}) + \rho_I(\xi,\bar{\xi}) w + \rho_I^2\beta_I(\xi,\bar{\xi})w^2 + \rho_I^3\gamma_I(\xi,\bar{\xi}) + \ldots \,.
\ee
This gives
\be
[\del_w h_I]^{-1} = \frac{1}{\rho_I} - 2 \beta_I w + (4\beta_I^2\rho_I - 3\gamma_I)w^2 + \ldots
\ee
from which one easily obtains the needed (needed for the ghost dilaton computation since the ghost dilaton only has modes $c_1$ and $c_{-1}$) $b$-ghost insertion coefficients
\bea
B_{-1}^I &=& \frac{1}{\rho_4} \delta_{I 4} \label{delta} \\
C_{-1}^I &=& 0 \label{zeroo} \\
B_{1}^I &=& \rho_I \del_\xi \beta_I + \rho_4 (4\beta_4^2 - 3 \gamma_4)\delta_{I 4} \\
C_{1}^I &=& \rho_I \del_{\bar{\xi}} \beta_I
\eea
where we took $z_1(\xi,\bar{\xi}) = 0$, $z_2(\xi,\bar{\xi}) = \infty$, $z_3(\xi,\bar{\xi}) = 1$ and $z_4(\xi,\bar{\xi}) = \xi$ and we do not insist on a symmetric vertex yet.
\subsection{Calculation of the integrand for $\langle T,T,T,D\rangle$}
\label{TTTDIntegrand}
The three fixed $T$ insertions are fixed only containing ghost factors of $c_1\bar{c}_1$ so that $\mathcal{B}$ and $\mathcal{B}^\star$ annihilates them thanks to the delta function in (\ref{delta}), giving
\be
\mathcal{B} = \sum_{m=-1}^\infty (B_m^{(4)} b_m^{(4)} + \overline{C}_m^{(4)} \bar{b}_m^{(4)}), \, \, \mathcal{B}^\star = \sum_{m=-1}^\infty (\overline{B}_m^{(4)} \bar{b}_m^{(4)} + C_m^{(4)} b_m^{(4)}) \,.
\ee
So that we can explicitly write
\be
\mathcal{B}\mathcal{B}^\star D = (B_{-1}^{(4)}b_{-1}^{(4)} + B_1^{(4)} b_1^{(4)} + \overline{C}_{1}^{(4)} \bar{b}_{1}^{(4)})(\overline{B}_{-1}^{(4)} \bar{b}_{-1}^{(4)} + \overline{B}_1^{(4)} \bar{b}_1^{(4)} + C_{1}^{(4)} b_{1}^{(4)})(c_1c_{-1} - \bar{c}_1\bar{c}_{-1})^{(4)}\ket{0}
\ee
where we used (\ref{zeroo}). Now using $\acomm{b_n}{c_m}=\delta_{n+m}$ and $b_n\ket{0} = 0$ for $n \geq {-1}$, we obtain
\be
\mathcal{B}\mathcal{B}^\star D = (-B_{-1}^{(4)} C_1^{(4)} - \overline{C}_1^{(4)}\overline{B}_{-1}^{(4)})\ket{0} \,.
\ee
Since $\mathcal{B}\mathcal{B}^\star D$ has only a component in the direction of $\ket{0}$, one obtains a $\langle T,T,T\rangle$ under the integrand. The local coordinates in this three-vertex can in principle still depend on $\xi$, but it turns out that one can factor out the $\langle T,T,T\rangle$ from the integral thanks to the antighost insertions at punctures $I = 1,2,3$ carrying modes $b_n$ with $n \geq 0$, which annihilate the $\ket{0}$. The rest is encoded in the form to integrate over (including a minus sign that comes from $ - \frac{1}{2\pi i}$ resulting in an overall minus in the definition of $\omega_D$ compared to Bergman and Zwiebach \cite{Bergman:1994qq})
\be
\omega_D = - d\xi \wedge d\bar{\xi} (-B_{-1}^{(4)} C_1^{(4)} - \overline{C}_1^{(4)}\overline{B}_{-1}^{(4)}) = d \xi \wedge d\bar{\xi} (\del_{\bar{\xi}} \beta_4 + \del_{\xi} \bar{\beta}_4) \,.
\ee
\subsection{Calculation of the integrand for $\langle T,T,D,D\rangle$}
\label{TTDDDer}
Now we examine $\mathcal{B}\mathcal{B}^\star (c_1c_{-1} - \bar{c}_1\bar{c}_{-1})^{(3)}(c_1c_{-1} - \bar{c}_1\bar{c}_{-1})^{(4)}\ket{0}$, where we again dropped the $c_1\bar{c}_1$ from the other fixed punctures (note that we cannot drop the ghost dilaton at puncture $3$ thanks to its nontrivial ghost structure despite the puncture being fixed). We then get
\bea
	& \mathcal{B}\mathcal{B}^\star (c_1c_{-1} - \bar{c}_1\bar{c}_{-1})^{(3)}(c_1c_{-1} - \bar{c}_1\bar{c}_{-1})^{(4)}\ket{0}  =  \nonumber \\ &(B_1^{(3)} b_1^{(3)} + \overline{C}_{1}^{(3)} \bar{b}_{1}^{(3)} + B_{-1}^{(4)}b_{-1}^{(4)} + B_1^{(4)} b_1^{(4)} + \overline{C}_{1}^{(4)} \bar{b}_{1}^{(4)}) \nonumber \\ &( \overline{B}_1^{(3)} \bar{b}_1^{(3)} + C_{1}^{(3)} b_{1}^{(3)} + \overline{B}_{-1}^{(4)} \bar{b}_{-1}^{(4)} + \overline{B}_1^{(4)} \bar{b}_1^{(4)} + C_{1}^{(4)} b_{1}^{(4)}) \nonumber \\ & \biggl[ - (c_1 c_{-1})^{(3)}(\bar{c}_1\bar{c}_{-1})^{(4)} - (\bar{c}_1\bar{c}_{-1})^{(3)} (c_1c_{-1})^{(4)}\biggr]\ket{0} = \nonumber \\ & \biggl[(C_1^{(3)}\overline{C}_1^{(4)}-B_1^{(3)}\overline{B}_1^{(4)})c_1^{(3)}\bar{c}_1^{(4)} + (\overline{C}_1^{(3)} C_1^{(4)}-\overline{B}_1^{(3)} B_1^{(4)}) \bar{c}_1^{(3)}c_1^{(4)} +  B_1^{(3)}\overline{B}_{-1}^{(4)} c_1^{(3)} \bar{c}_{-1}^{(4)} + \overline{B}_1^{(3)} B_{-1}^{(4)} \bar{c}_1^{(3)} c_{-1}^{(4)}\biggr]\ket{0} \nonumber \\
\eea
where we dropped factors whose ghost number doesn't add up to $(1,1)$ keeping in mind that we insert two $T$s at the other punctures. This means that we need to compute the ghost correlators \\ $\langle (c_1\bar{c}_1)^{(1)}, (c_1\bar{c}_1)^{(2)}, c_1^{(3)}, \bar{c}_1^{(4)}\rangle$ and $\langle (c_1\bar{c}_1)^{(1)}, (c_1\bar{c}_1)^{(2)}, c_1^{(3)}, \bar{c}_{-1}^{(4)}\rangle$. We note that the insertion at infinity is accompanied by a damping factor from the conformal transformation giving $\lim_{z\to \infty} \frac{1}{\abs{z}^4} c \bar{c}(z)$. This gives
\be
\langle (c_1\bar{c}_1)^{(1)}, (c_1\bar{c}_1)^{(2)}, c_1^{(3)}, \bar{c}_1^{(4)}\rangle = \frac{1}{\rho_1^2\rho_2^2\rho_3\rho_4}\lim_{z\to \infty} \frac{1}{\abs{z}^4}\langle c \bar{c}(0) c\bar{c}(z) c(1) \bar{c}(\xi) \rangle = \frac{\bar{\xi}}{\rho_1^2\rho_2^2\rho_3\rho_4},
\ee
where we used the normalisation
\be
\langle c \bar{c}(z_1) c\bar{c}(z_2) c\bar{c}(z_3)\rangle = \langle c(z_1)c(z_2)c(z_3) \rangle \langle \bar{c}(z_1)\bar{c}(z_2)\bar{c}(z_3)\rangle
\ee
with
\be
\langle c(z_1)c(z_2)c(z_3)\rangle = (z_1-z_2)(z_1-z_3)(z_2-z_3) \,.
\ee
For the second correlator, we note the transformation property of the operator $\frac{1}{2}\bar{\del}^2\bar{c}$ being the image of $\bar{c}_{-1}$ under the state operator mapping
\be
\frac{1}{2}\bar{\del}^2\bar{c}(w) = \rho_I\biggr(\frac{1}{2} \bar{\del}^2\bar{c}(\bar{z}_I) - \bar{\beta}_I\bar{\del}\bar{c}(\bar{z}_I) + \frac{\bar{\epsilon}_I}{2} \bar{c}(\bar{z}_I)\biggr)
\ee
with
\be
\epsilon_I = 8 \beta_I^2 - 6 \gamma_I \,.
\ee
A simple computation gives
\be
\langle (c_1\bar{c}_1)^{(1)}, (c_1\bar{c}_1)^{(2)}, c_1^{(3)}, \bar{c}_{-1}^{(4)}\rangle = \frac{\rho_4}{\rho_1^2\rho_2^2\rho_3} \biggr(\frac{1}{2}\epsilon_4 \bar{\xi} - \beta_4\biggr) \,.
\ee
Next we have
\bea
C_1^{(3)}\overline{C}_1^{(4)}-B_1^{(3)}\overline{B}_1^{(4)} &=& \rho_3\rho_4(\del_{\bar{\xi}}\beta_3 \del_{\xi}\beta_4 - \del_{\xi}\bar{\beta}_3 \del_{\xi}\bar{\beta}_3 - \frac{1}{2}\epsilon_4 \del_\xi \beta_3) \\
B_1^{(3)}\overline{B}_{-1}^{(4)} &=& \frac{\rho_3}{\rho_4}\del_\xi \beta_3 \,.
\eea
So that
\bea
                                 &(C_1^{(3)}\overline{C}_1^{(4)}-B_1^{(3)}\overline{B}_1^{(4)} ) \langle (c_1\bar{c}_1)^{(1)}, (c_1\bar{c}_1)^{(2)}, c_1^{(3)}, \bar{c}_1^{(4)}\rangle  + B_1^{(3)}\overline{B}_{-1}^{(4)} \langle (c_1\bar{c}_1)^{(1)}, (c_1\bar{c}_1)^{(2)}, c_1^{(3)}, \bar{c}_{-1}^{(4)}\rangle = \nonumber \\ & \frac{1}{\rho_1^2\rho_2^2}\biggr(\del_{\bar{\xi}}\beta_3\del_\xi(\bar{\xi}\bar{\beta}_4)-\del_\xi \beta_3\bar{\del}(\bar{\xi}\bar{\beta}_4)\biggr) \,.
\eea
From matter we have the two-point function
\be
\langle V^{(1)} V^{(2)}\rangle = \rho_1^{2h}\rho_2^{2h} = \rho_1^{2}\rho_2^{2} + O(y)
\ee
and taking into account the $\star$-conjugate present after the action of $\mathcal{B}\mathcal{B}^\star$, we finally obtain the integrand $-2 d\xi \wedge d\bar{\xi} \Re \biggr(\del_{\bar{\xi}}\beta_3\del_\xi(\bar{\xi}\bar{\beta}_4)-\del_\xi \beta_3\bar{\del}(\bar{\xi}\bar{\beta}_4)\biggr) + O(y) $ with the minus again coming from $\frac{i}{2\pi} = -\frac{1}{2\pi i}$. This integrand can be written as $d\Omega$ with
\be
\Omega = - \Re \biggr([(\del_\xi \beta_3)\bar{\xi}\bar{\beta}_4 -\beta_3\del_\xi(\bar{\xi}\bar{\beta}_4)] d\xi + [(\del_{\bar{\xi}}\beta_3)\bar{\xi}\bar{\beta}_4-\beta_3\del_{\bar{\xi}}(\bar{\xi}\bar{\beta}_4)]d\bar{\xi}\biggr) + O(y) \,.
\ee

\sectiono{Warming up with stubbed OSFT}
\label{osft}
In this appendix, we redo the analysis of this paper in stubbed OSFT, essentially reproducing \cite{Scheinpflug:2023osi}. We do this because the geometry is more transparent in this case (the disc is simpler than the sphere so that the calculations of (\ref{reg}) and (\ref{lens}) become less subtle) and because there is no ghost dilaton in OSFT so that we know that in this case our results hold in the full theory.

\begin{figure}[h!]
	\centering
	\includegraphics[width=7cm]{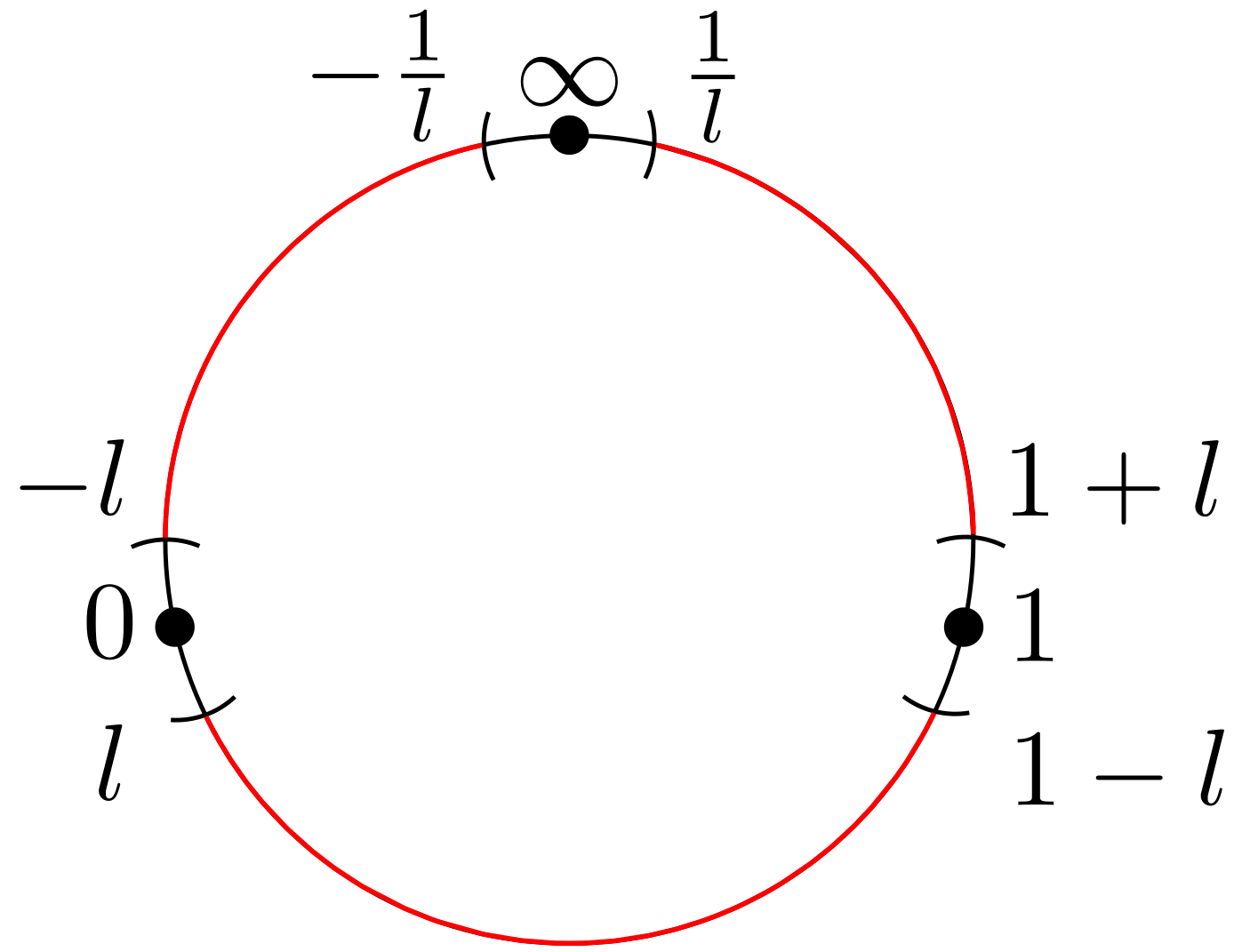}
	\caption{The vertex region of the four-punctured disc is depicted in red}
	\label{disc}
\end{figure}

	We would like to compute a four-tachyon amplitude in stubbed OSFT. In \cite{Scheinpflug:2023osi} we've shown that $\langle T * T,  \frac{b_0}{L_0} \bar{P} (T * T)\rangle$ with $\bar{P}$ projecting out $T$ covers $\frac{1}{6}$ of moduli space (we enountered an integral from $0$ to $\frac{1}{2}$, see figure \ref{disc} and also note that there are 6 ways to put four vertex operators on the boundary of a disc) so that in the stubbed theory the full four-tachyon amplitude of $T = t c_1 V\ket{0}$ is
\be
t^4 \mathcal{A}_{TTTT} = 6 \langle T \star T,  \frac{b_0}{L_0} \bar{P} (T \star T)\rangle + \langle T,T,T,T\rangle
\ee
where now $\star$ is a stubbed Witten's product and $\langle T,T,T,T\rangle$ is a contact interaction filling the rest of moduli space (when we introduce stubs, the propagator region shrinks, see \cite{Schnabl:2023dbv,Erbin:2023hcs}).

If we have the fusion $V\cross V = 1 + V$, then in the infinite stub limit only the identity propagates (remember that we have a $\bar{P}$) so that
\be
6 \langle T \star T,  \frac{b_0}{L_0} \bar{P} (T \star T)\rangle = -6 g_0 \abs{\lambda h'_3(1)}^2 + O(y)
\label{feyn}
\ee
where we used that $V$ is nearly marginal meaning that we can expand in $y \equiv 1-h$ with $h$ being its weight and we denote the $g$-function of the initial background as $g_0$. For the details on the local coordinates, see subsection \ref{choicev} and make the very simple translation of the formulas from the sphere to the disc.

The vertex region integrand is easily (by cutting out neighborhoods around punctures, see figure \ref{disc}) seen to be
\be
\frac{1}{g_0 t^4}\langle T,T,T,T\rangle = \, \int\displaylimits_\mathcal{R}  d\xi \, \, \bra{V} V(1)V(\xi) \ket{V} \label{rewr}
\ee
where we denote by $l \equiv \frac{1}{\abs{\lambda h'_3(1)}^2}$ a cutoff naturally provided by SFT (small in the large stub limit $\lambda \to \infty$) and
\be
\int\displaylimits_\mathcal{R} d\xi = \int\displaylimits_{-\frac{1}{l}}^{-l}  d\xi + \int\displaylimits_{l}^{1-l} d\xi + \int\displaylimits_{1+l}^{\frac{1}{l}} d\xi \,.
\ee

After identifying the channels that contribute to divergences by
\bea
\bra{V} V(1) V(\xi) \ket{V} = (\xi^{-2} + \ldots) + \threept^2 (\xi^{-1} + \ldots) + O(y),
\eea
we can split (\ref{rewr}) into a finite contribution where we integrate over the entire line $\rr$ (an integral over a bounded function over $\mathcal{R}$ is the same upon $\mathcal{R} \to \rr$ when we take the large stub limit) and a divergent contribution. By subtracting and adding the divergent part, we have

\bea
\frac{1}{g_0t^4} \langle T,T,T,T\rangle &=& \int\displaylimits_\rr d\xi \, \, \biggr[ \bra{V} V(1) V(\xi) \ket{V} - \frac{1}{2}\biggr(\frac{1}{\xi^2(\xi-1)^2} + \frac{(\xi-1)^2}{\xi^2} + \frac{\xi^2}{(\xi-1)^2} \nonumber \biggr)  \nonumber \\ && \hspace{1 cm} \nonumber - \frac{\threept^2}{2}\biggr( \frac{1}{\abs{\xi}\abs{\xi-1}} + \frac{1}{\abs{\xi}} + \frac{1}{\abs{\xi-1}}\biggr) \biggr] +
				     \\ &&  \int\displaylimits_\mathcal{R} d\xi \, \, \biggr[ \frac{1}{2}\biggr(\frac{1}{\xi^2(\xi-1)^2} + \frac{(\xi-1)^2}{\xi^2} + \frac{\xi^2}{(\xi-1)^2}\biggr) \nonumber \\ && \hspace{1 cm} + \frac{\threept^2}{2}\biggr( \frac{1}{\abs{\xi}\abs{\xi-1}} + \frac{1}{\abs{\xi}} + \frac{1}{\abs{\xi-1}}\biggr) \biggr] + O(y)
\eea
where the subtraction was derived by having the correct asymptotics around punctures and being $Sl(2,\rr)$ symmetric. Note that one can write one of the subtractions in a more intuitive way
\be
\frac{1}{2}\biggr(\frac{1}{\xi^2(\xi-1)^2} + \frac{(\xi-1)^2}{\xi^2} + \frac{\xi^2}{(\xi-1)^2} \nonumber \biggr)  = 1 + \frac{1}{(\xi-1)^2} + \frac{1}{\xi^2}
\ee
but that such a simplification does not occur in the closed string.
Carrying out the $\mathcal{R}$ integral and expanding it in the stub gives
$6 \abs{\lambda h'_3(1)}^2 + 12 \threept^2\ln\abs{\lambda h'_3(1)}$, which together with the Feynman contribution (\ref{feyn}) results in
\bea
\mathcal{A}_{TTTT} &=& -6g_0 \abs{\lambda h'_3(1)}^2 + 6g_0 \abs{\lambda h'_3(1)}^2 + 12 g_0 \threept^2\ln\abs{\lambda h'_3(1)} + \mathcal{A}_{TTTT}^{f.p.} + O(y) \nonumber \\ &=& 12 g_0 \threept^2\ln\abs{\lambda h'_3(1)} + \mathcal{A}_{TTTT}^{f.p.} \label{plug}
\eea
with
\bea
\mathcal{A}_{TTTT}^{f.p.} &=& g_0\int\displaylimits_\rr d\xi \, \, \biggr[ \bra{V} V(1) V(\xi) \ket{V} -\biggr(1 + \frac{1}{(\xi-1)^2} + \frac{1}{\xi^2} \nonumber \biggr)  \nonumber \\ && \hspace{1 cm}  - \frac{\threept^2}{2}\biggr( \frac{1}{\abs{\xi}\abs{\xi-1}} + \frac{1}{\abs{\xi}} + \frac{1}{\abs{\xi-1}}\biggr) \biggr] \label{plugg} \,.
				 \eea

By using that the shift in the $g$-function can be calculated in OSFT by $\Delta g = -2\pi^2 S$ where $S$ is the value of the OSFT on-shell action , we've shown in \cite{Scheinpflug:2023osi} that

\be
\frac{\Delta g}{g_0} = -\frac{\pi^2}{3}\biggr(\frac{y^3}{\threept^2}  + \biggr(\frac{1}{g_0}\frac{\mathcal{A}_{TTTT}}{2\threept^4} -  \frac{6}{\threept^2} \ln \abs{\lambda h'_3(1)}\biggr)y^4\biggr) + O(y^5).
\ee
Plugging in (\ref{plug}) then finally gives
\be
\frac{\Delta g}{g_0} = -\frac{\pi^2}{3}\biggr(\frac{y^3}{\threept^2}  + \frac{1}{g_0}\frac{\mathcal{A}_{TTTT}^{f.p.}}{2\threept^4}y^4\biggr) + O(y^5) \label{gres},
\ee
which goes beyond the classic result of Affleck and Ludwig \cite{Affleck:1991tk,Ludwig:1994nf}.
Observe the similarity of (\ref{plugg}) to (\ref{fourtachyon}) and of (\ref{gres}) to (\ref{cecko}). This manifestly passes a trivial check that $\mathcal{A}_{TTTT}^{f.p.} = 0$ for a $c=1$ free boson where
\be
\bra{V} V(1) V(\xi) \ket{V} = 1 + \frac{1}{(\xi-1)^2} + \frac{1}{\xi^2}
\ee
and we also tested it on a so-called exotic solution in a $c=2$ free boson \cite{Exotic}. It is remarkable how much simpler the derivation in the infinitely stubbed theory is when compared to the flattenization used by \cite{Scheinpflug:2023osi}.
\printbibliography[]

\end{document}